\documentclass[10pt,letterpaper]{article}
\usepackage{opex3}

 \usepackage{subfigure,amsmath,amssymb}

\begin{document}

\title{Improved detection of atmospheric turbulence with SLODAR}

\author{Michael Goodwin, Charles Jenkins }

\address{Research School of Astronomy and Astrophysics, Australian National University, Cotter Road, Weston ACT 2611, Australia}

\author{ Andrew Lambert}

\address{School of Information Technology and Electrical Engineering,
Australian Defence Force Academy, University of New South Wales,
Canberra, ACT 2600, AUSTRALIA}

\email{mgoodwin@mso.anu.edu.au} 



\begin{abstract}
We discuss several improvements in the detection of atmospheric turbulence using SLOpe Detection And Ranging (SLODAR).  Frequently, SLODAR observations have shown strong ground-layer turbulence, which is beneficial to adaptive optics. We show that current methods which neglect atmospheric propagation effects can underestimate the strength of high altitude turbulence by up to $\sim30\%$. We show that mirror and dome seeing turbulence can be a significant fraction of measured ground-layer turbulence, some cases up to $\sim50\%$. We also demonstrate a novel technique to improve the nominal height resolution, by a factor of 3, called Generalized SLODAR. This can be applied when sampling high-altitude turbulence, where the nominal height resolution is the poorest, or for resolving details in the important ground-layer.\\
\end{abstract}



\section{Introduction}

The success of adaptive optics in astronomy has been demonstrated with  {\color{red} } 8-10m class telescopes. Performance has been reported with the 10m Keck II Telescope~\cite{vanDam}, the 8m Very Large Telescope~\cite{Rousset}, the 8.2m Subaru Telescope~\cite{Iye} and 8m Gemini North Telescope~\cite{Stoesz} and others. The performance of adaptive optics systems depends strongly on the characteristics of the atmospheric turbulence above the telescope~\cite{Hardy}. Information about the height distribution of the atmospheric turbulence in terms of its strength and velocity can be used to optimize adaptive optic models, and prove the case for future installations. Measurements of the atmospheric turbulence can reveal important parameters for adaptive optics, the coherence length, $r_0$, the coherence time, $\tau_0$, and the anisoplantanic angle, $\theta_0$. Other useful parameters include the outer scale, $L_0$, and the power law, $\beta$, of the power spectrum of spatial phase fluctuations (for Kolmogorov, $L_0=\infty$ and $\beta=11/3$).
Measurement of these parameters has been emphasized with the planned construction of Extremely Large Telescopes (ELT) and new adaptive optic technologies, such as Ground Layer Adaptive Optics (GLAO). This has led to numerous campaigns to characterize the atmospheric turbulence profile at current or proposed observatory sites, for example, the Cerro Tololo campaign~\cite{TokovininTravouillon}. At many sites a significant fraction of the turbulence has been found near the ground, which is favorable for GLAO. This is promising as GLAO relies on compensating the low altitude turbulence and providing a uniform partial correction over relatively wide-field of several arcminutes~[7-9]. 

Various techniques are used for turbulence ranging, including direct
sensing with microthermal sensors on towers~\cite{Pant} or balloons
~\cite{Azouit}, remote-sensing with acoustic scattering (SODAR) ~\cite{Travouillon}, or triangulation of scintillation (SCIDAR)~\cite{Vernin,Fuchs} or of image motion (SLODAR) ~\cite{Wilson,Butterley}. These techniques have reached a degree of maturity exhibiting reasonable agreement when used together in campaigns (\cite{TokovininTravouillon}, Cerro Tololo campaign). Each technique has its benefits and limitations in terms of cost, height resolution, height range, temporal resolution, ease of implementation and data reduction complexity.

\begin{figure} [htbp]
\centering
  \includegraphics[width=0.95\textwidth]{Goodwin_SLODAR_fig1.eps}
  \caption{Diagram illustrating the geometry of the SLODAR method for a N=4 system. $\theta$ is the double star angular separation. D is the diameter of the telescope pupil and $w$ the width of the sub-aperture of the Shack-Hartmann Wavefront Sensor (SHWFS) array. The centers of the altitude bins are given by $\Delta \delta h$ where $\delta h = w/\theta$. The ground-layer can be analyzed in higher-resolution by utilization of double stars having larger $\theta$. The diagram illustrates the notation to describe the SLODAR theoretical covariance impulse function,  $X_L (\Delta ,\delta i,\delta j)$ as well as the notation for Generalized SLODAR (see Section \ref{sec:verticalres}). }\label{fig:SLODAR_Geometry}
\end{figure}

The SLODAR technique, Fig.~\ref{fig:SLODAR_Geometry}, is based on Shack-Hartmann Wavefront Sensor (SHWFS) that measures the averaged local wavefront derivative or slope across the telescope pupil using an array, ($N \times N$), of square sub-apertures or lenslets. The 2-D spatial cross-covariance of the sub-aperture spot motions (or Z-tilts) of the double star components in each frame are averaged. Then a 1-D slice is taken along the double star separation axis and then inverted against the 1-D theoretical covariance impulse functions providing an estimate of the $C_N^2(h)dh$~\cite{Butterley}. The process of fitting covariance impulse functions allows the estimation of the outer scale, $L_0$, and the power law, $\beta$, of the power spectrum of spatial phase fluctuations. The process of obtaining $C_N^2(h)dh$ information from the observational data is further explained in Section~\ref{sec:verticalres}. The vertical resolution is uniform, given by $\delta h = w / \theta $ where $w$ is sub-aperture size, or lenslet size mapping to the telescope pupil. The highest sampled layer, $H_{max} = (N-1) \times \delta h \approx D / \theta $ , where N is number of sub-apertures across the telescope pupil, with the ground layer denoted as $h_{0} = 0$. The vertical resolution and maximum sample height are scaled by $\cos(\zeta)$, where $\zeta$ is the zenith distance.

The exposure times are typically 4~ms to 8~ms and directly proportional to sub-aperture size, related to the wind speed crossing timescales. The sub-aperture sizes are designed to be approximately equal or less than $r_0$, or ranging from 5~cm to 15~cm depending on the median seeing. Sensitivity to higher altitude turbulence is reduced because there are fewer longer baselines in the pupil, inability to freeze turbulence due to high wind speeds, turbulence strength is usually weak (compared to ground-layer) and covariance impulse response decreases with altitude due to propagations effects. This paper will attempt to lessen or mitigate the problems arising from these factors, particularly focussing on the propagation effects, removal of mirror and dome seeing turbulence and improving the height resolution.

With a scientific motivation to determine the statistical properties of the height distribution of turbulence at medium quality astronomical site, we have been pursuing an extensive SLODAR campaign to characterize the atmospheric turbulence at the Siding Spring Observatory (SSO). Data taken consists of 7x7 SLODAR instrument on the ANU (Australian National University) 24" (1"=2.54cm) telescope and 17x17 SLODAR instrument ANU 40" telescope. A portion of data taken with the ANU 40" telescope implemented real-time data processing at 200 fps to improve the observational data quality and reduce storage requirements (no need to store raw camera frames). The ANU 17x17 SLODAR instrument has obtained $\delta h = 75~\rm{m}$ and $H_{max} = 1200~\rm{m}$ when observing $\delta~\rm{Apodis}$ having separation, $\theta = 102.9"$ with a zenith distance, $\zeta=50^\circ$. For high altitude sampling, we have obtained $\delta h = 2400~\rm{m}$ and $H_{max} = 40800~\rm{m}$ when observing $\alpha~\rm{Crucis}$ having separation, $\theta = 4.1"$ with a zenith distance, $\zeta=35^\circ$.

We have discovered a tendency for the usual implementation of SLODAR to underestimate the strength of high turbulent layers. This was later confirmed in SLODAR numerical simulations involving phase screens with Fresnel and Geometrical propagation techniques. To examine in further detail, theoretical calculations were implemented for the covariance impulse functions~\cite{Butterley} using modified phase power spectrum to model Fresnel propagation. We describe the propagation effects on SLODAR in Section~\ref{sec:propagation}.

Also discovered was that the majority of the turbulence profiles were dominated by the ground-layer or zero altitude contribution, $h_{0}=0$, found in part to be caused by strong mirror and dome seeing turbulence. By applying a high pass filter with cut-off of 1-2 Hz to the temporal centroid data streams, it was possible to remove the mirror and dome seeing turbulence from the ground-layer measurement. However, at this stage we point out that the ground-layer at Siding Spring dominates the seeing, particularly so on nights with poor seeing. We describe the process of removing dome and mirror seeing turbulence from SLODAR data in Section~\ref{sec:mirrorturb}.

In order to obtain improved vertical resolution, we have modified the instrument to optically move the zero height analysis plane from the telescope pupil upwards to fractional heights of the nominal height resolution, $\delta h$. We call this technique Generalized SLODAR and we report on the methodology, numerical simulations and observational results in Section~\ref{sec:verticalres}.

A full report on the Siding Spring SLODAR campaign, which now covers one week per season for 18 months, will be forthcoming in a later publication.

\section{Propagation Effects of High-Altitude Layers}
\label{sec:propagation}

The retrieval of the turbulence profile was initially described by the method outlined by Wilson~\cite{Wilson}, as the spatial cross-correlation of the centroid data from star A and star B de-convolved with the spatial auto-correlation of the centroid data from star A, see Fig.~\ref{fig:SLODAR_Geometry}. The global X and Y tilts of the double star components, A and B, are subtracted from the centroid data, to remove any telescope tracking errors. The method assumes the auto-covariance ("auto-correlation" by Wilson~\cite{Wilson}) is the spatial invariant impulse response of a thin layer for all sampling heights. This assumption simplifies the data reduction but neglects the effects of the global tilt subtraction. Later, Butterley et al.~\cite{Butterley} included the effects of global tilt subtraction by calculation of the theoretical covariance impulse functions, based on the power spectrum of phase fluctuations of a thin turbulent layer located at each sampling height. For Kolmogorov turbulence, Butterley et al.~\cite{Butterley} show that global tilt subtraction adds tilt-anisoplanatism which could under estimate the strength of the highest sampling altitude by up to 20\%. The tilt-anisoplanatism is caused by the separation of the projected telescope pupils of star A and B onto the turbulent layer, see Fig.~\ref{fig:SLODAR_Geometry}.

Butterley et. al.~\cite{Butterley} calculations do not take into account propagation effects of the high altitude turbulent layers to the telescope pupil where SLODAR analysis is performed. Propagation may be included by using a modified input power spectrum of spatial phase fluctuations, derived by Roddier~\cite{Roddier} as

\begin{equation}
P^0_{\phi}(f) = P_{\phi}(f) \cos^2(\pi \lambda h f^2)
\label{eqn:powerspec}
\end{equation}

in which $P_{\phi}(f)$ is the phase power spectrum with no
propagation, in other words as the wavefront leaves the layer at
height $h$.  $P^0_{\phi}(f)$ is then the power spectrum at the ground, $h=0$. The nulls of the propagated power spectrum, $P^0_{\phi}(f_n)=0$ occurs at spatial frequencies, $f_n = [(n+0.5)^{1/2}]/r_f$, for integer $n=0,1,...$, and where $r_f=(\lambda h)^{1/2}$ denotes the layer's Fresnel length. This results in a measurable decrease, in the variance of the image motion across sub-aperture with size comparable to the Fresnel length. The effect is increased by the removal of global tilt in the reduction process, as this eliminates low spatial frequencies.

Therefore, propagation effects are most important for small sub-apertures such as those employed in SLODAR. For example, a turbulent layer at $h=15~\rm{km}$ has a Fresnel length of 8.6~cm at a wavelength of 0.5 microns. In our site testing observations at Siding Spring, we used small (5.8~cm on the 40" and 8.5~cm on the 24" telescopes) sub-apertures for SLODAR because the seeing is often poor, with $r_0$ about 8~cm in median seeing. These sub-apertures sizes are similar to the 5~cm sub-apertures used by the European Southern Observatory (ESO) portable SLODAR system using a 40~cm telescope~\cite{Wilson,Butterley}.

\begin{figure}[]
  \begin{center}
    \mbox{
      \subfigure[]{\scalebox{1.0}{\includegraphics[width=0.45\textwidth]{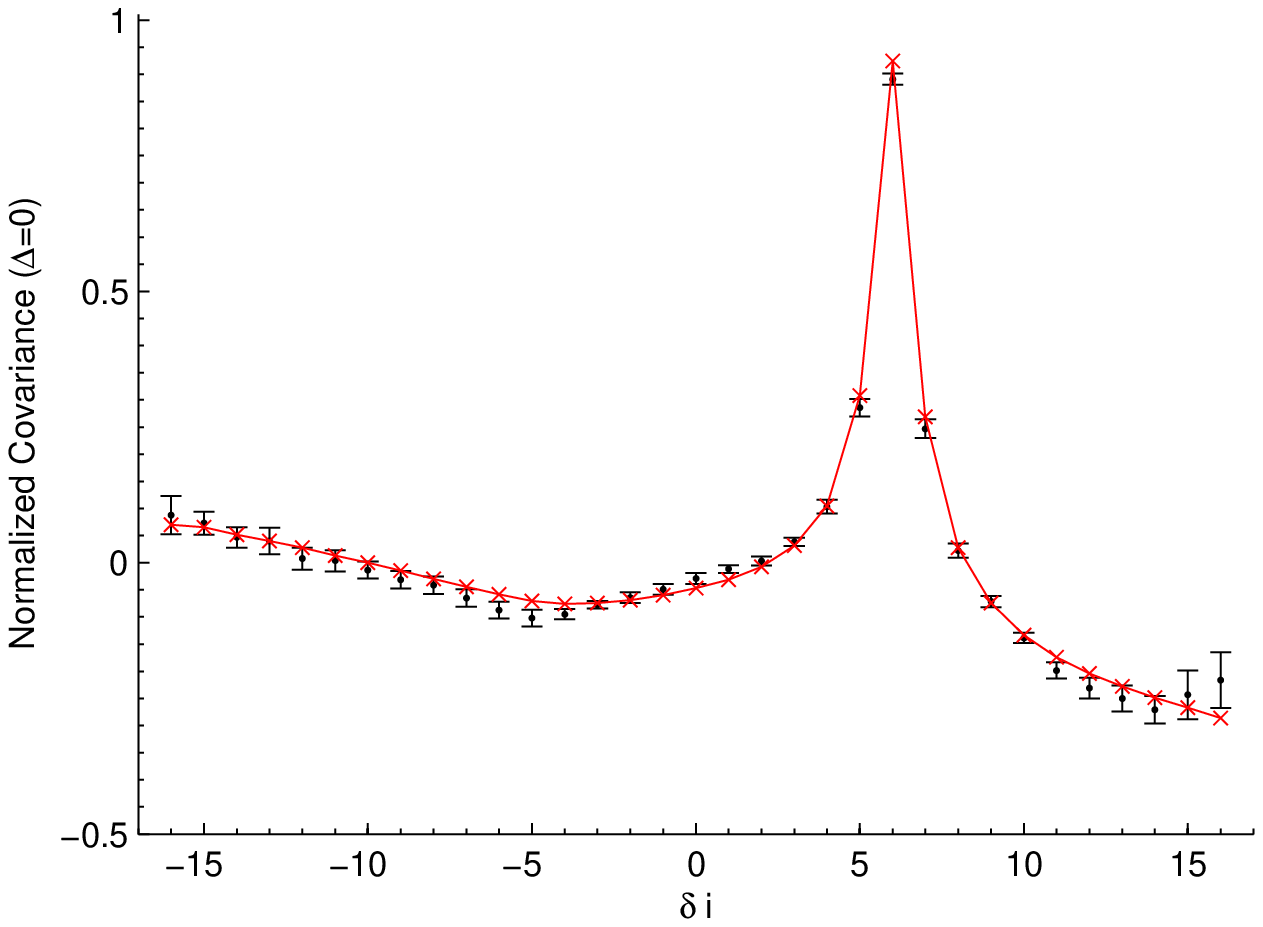}}} \quad
      \subfigure[]{\scalebox{1.0}{\includegraphics[width=0.45\textwidth]{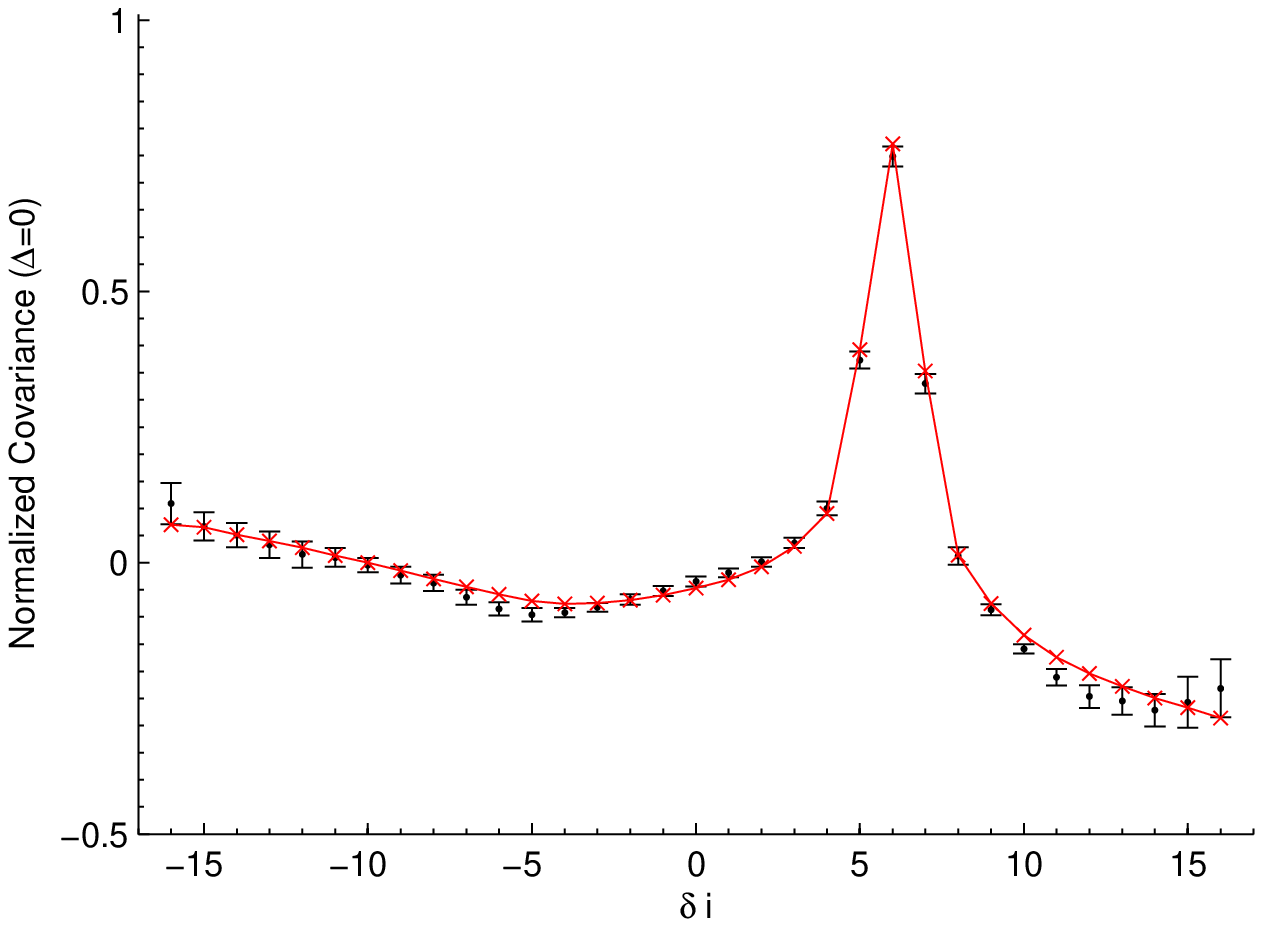}}}
      }
    \mbox{
      \subfigure[]{\scalebox{1.0}{\includegraphics[width=0.45\textwidth]{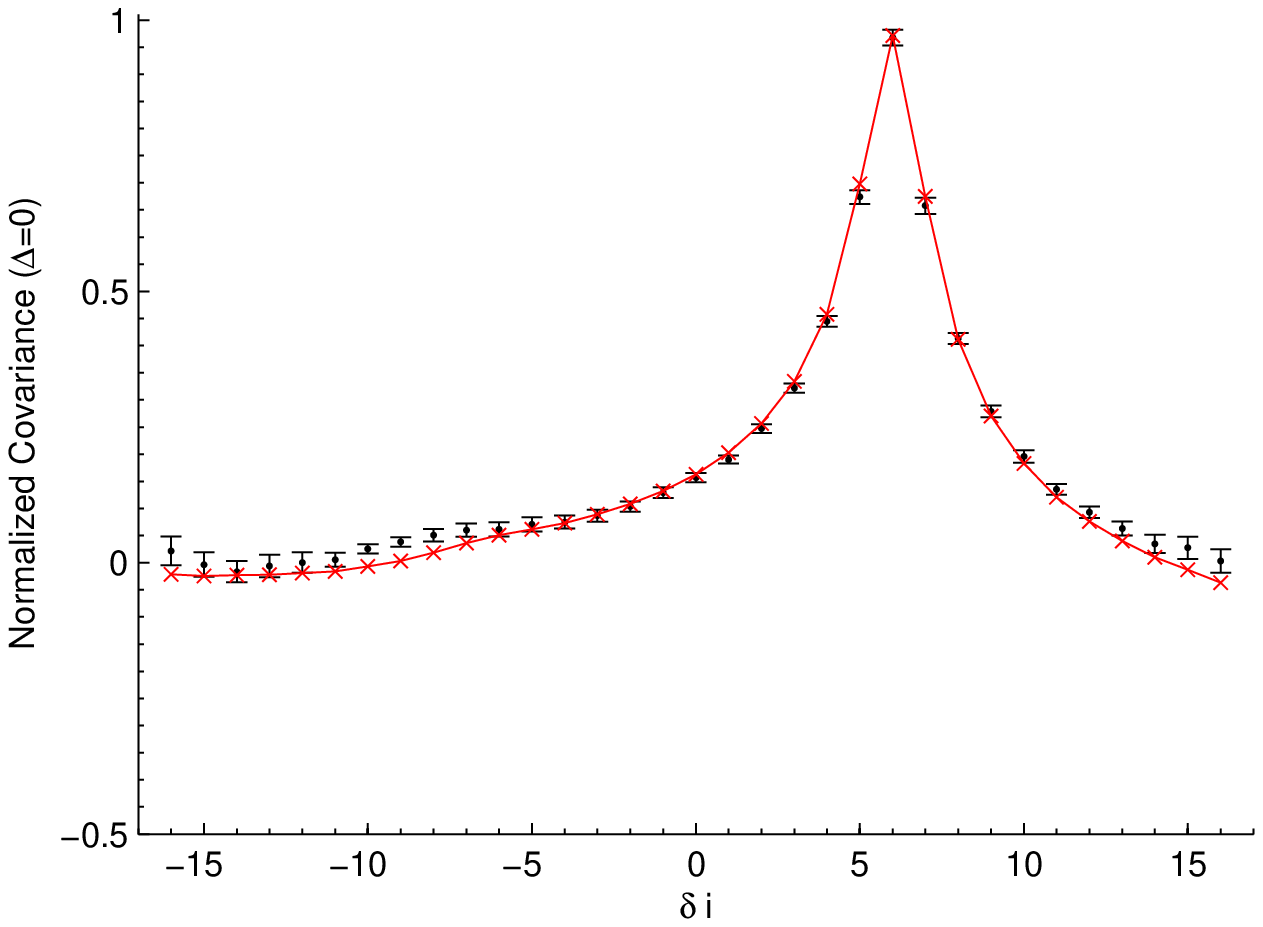}}} \quad
      \subfigure[]{\scalebox{1.0}{\includegraphics[width=0.45\textwidth]{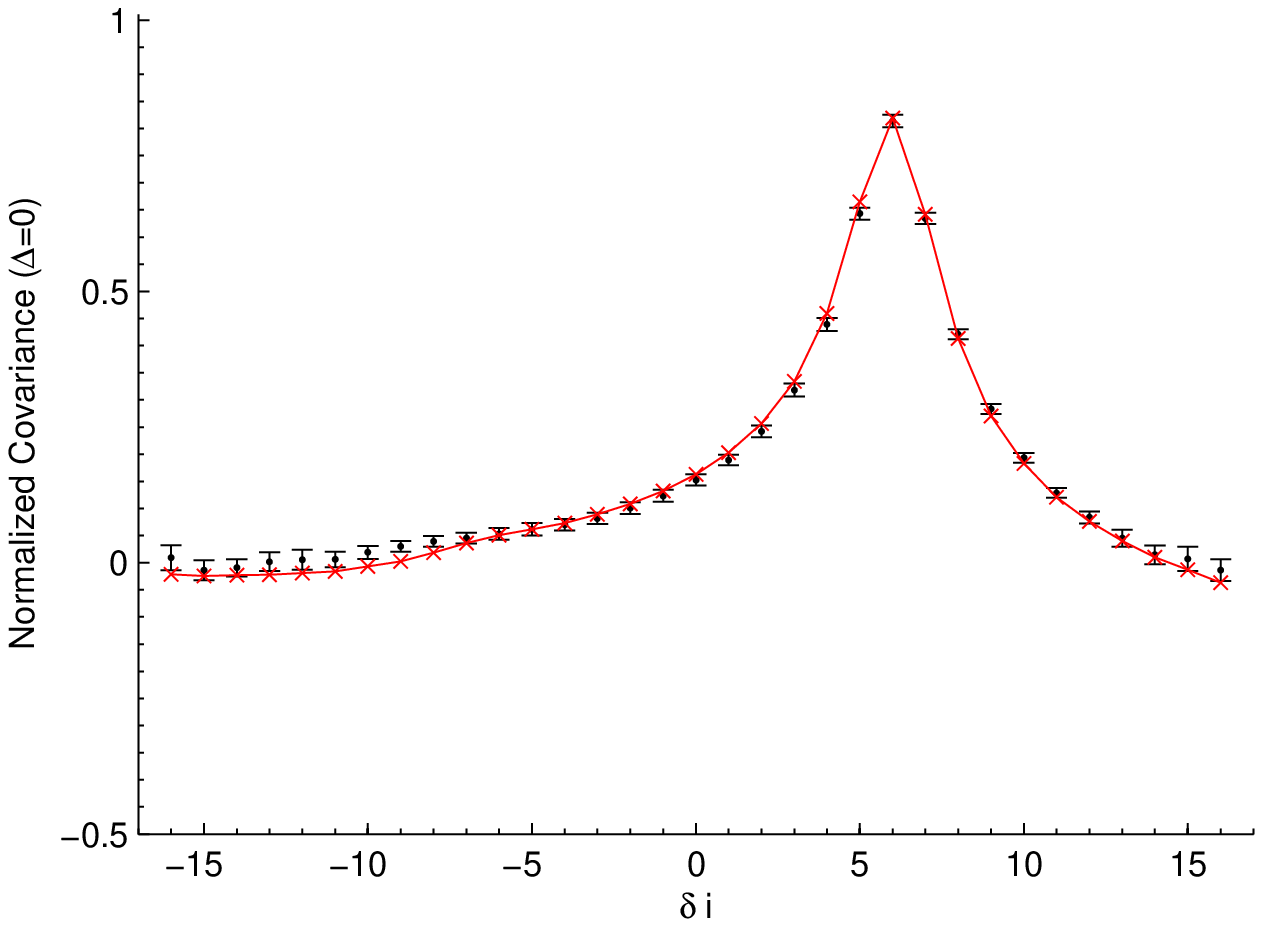}}}
      }
    \caption{Comparison of propagation effects on the covariance response function using numerical simulations (black dots with error bars) using Fresnel propagation and with theoretical values (red line with crosses) using the modified power spectrum (Eq.~\ref{eqn:powerspec}). The covariance plots are (a) longitudinal no propagation, (b) longitudinal with propagation, (c) transverse no propagation and (d) transverse with propagation. The longitudinal direction refers to direction parallel to double star separation axis, aligned along the x-direction of the SHWFS. The transverse direction refers to direction perpendicular to double star separation axis, aligned along the y-direction of the SHWFS. The comparisons are for single turbulent with $\Delta=6$ or height $H=7709~\rm{m}$ and normalized by their respective $\Delta=0$ or height $H=0~\rm{m}$ functions (i.e. no fitting involved). Plots (b) and (d) show that propagation effects decrease the peak covariance value ($\delta i = 6$) by $\sim20\%$ compared to plots (a) and (c).}
    \label{fig:SLODAR_Numerical_Propagation}
  \end{center}
\end{figure}

In Fig.~\ref{fig:SLODAR_Numerical_Propagation} we compare the effects of propagation for a turbulent layer at $H=7709~\rm{m}$, or pupil separation, $\Delta=6$, projected onto H, where $\Delta=H\theta/w$. The target double star referenced in calculations is $\alpha~\rm{Cen}$ with separation, $\theta = 9.44"$. The plots show the corresponding covariance impulse function $\Delta=6$ calculated with numerical simulations involving 300 phase screens using Fresnel propagation (only for propagation effects) and the calculated theoretical covariance impulse response using the methodology outlined by Butterley et al.~\cite{Butterley} with the modified power spectrum of phase fluctuations (Eq.~\ref{eqn:powerspec}) (only for propagation effects). The numerical and theoretical models for the results in Fig.~\ref{fig:SLODAR_Numerical_Propagation} reference the pupil geometry of the ANU 17x17 SLODAR instrument on the ANU 40" telescope.  It is evident that propagation decreases the covariance response peak by $\sim20\%$ causing a broadening effect with increasing height. The results of Fig.~\ref{fig:SLODAR_Numerical_Propagation} show an excellent agreement between numerical and theoretical results, hence validating the use of Eq.~\ref{eqn:powerspec} in theoretical calculations and analysis.

\begin{figure}[htbp]
  \begin{center}
    \mbox{
      \subfigure[]{\scalebox{1.0}{\includegraphics[width=0.45\textwidth]{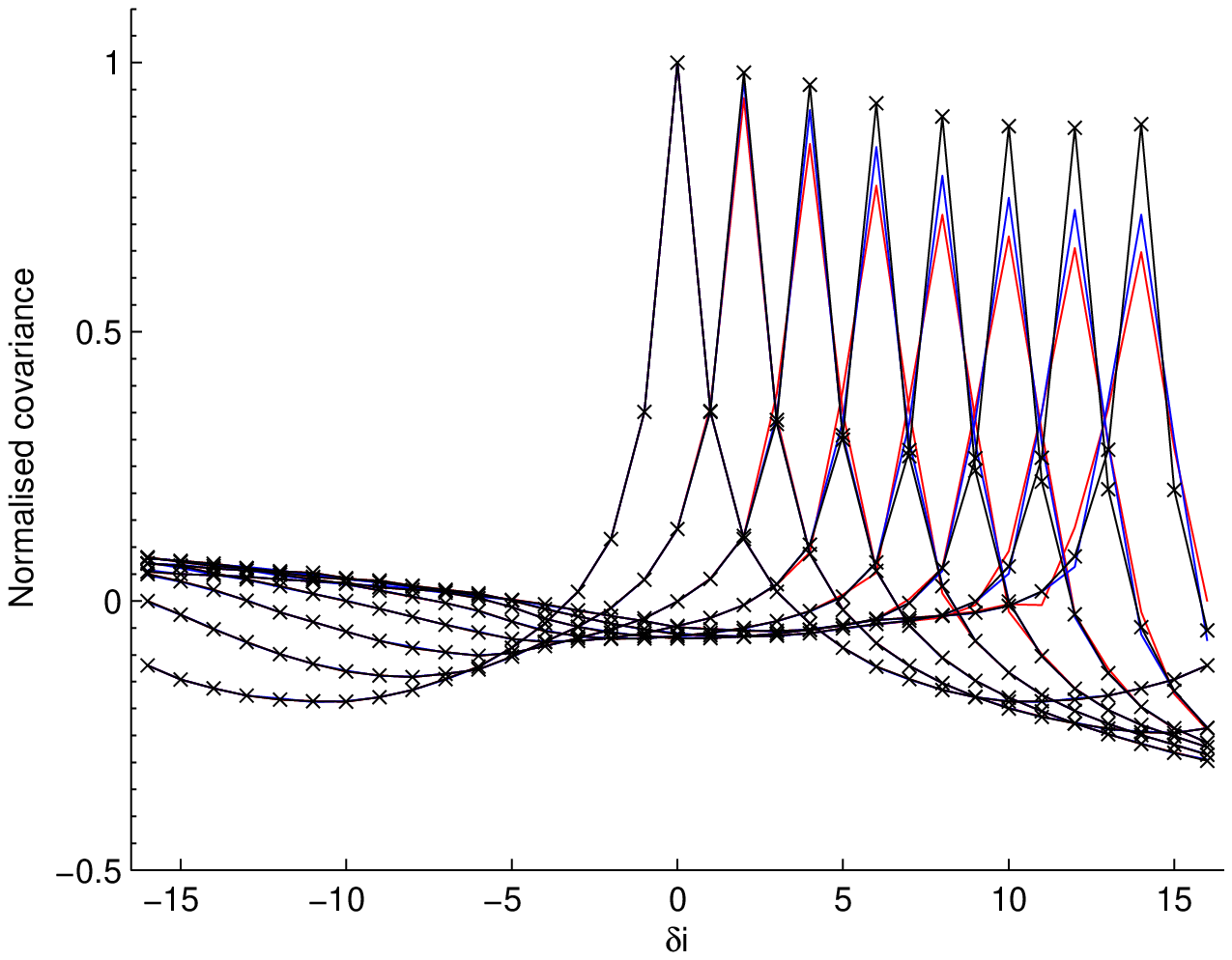}}} \quad
      \subfigure[]{\scalebox{1.0}{\includegraphics[width=0.45\textwidth]{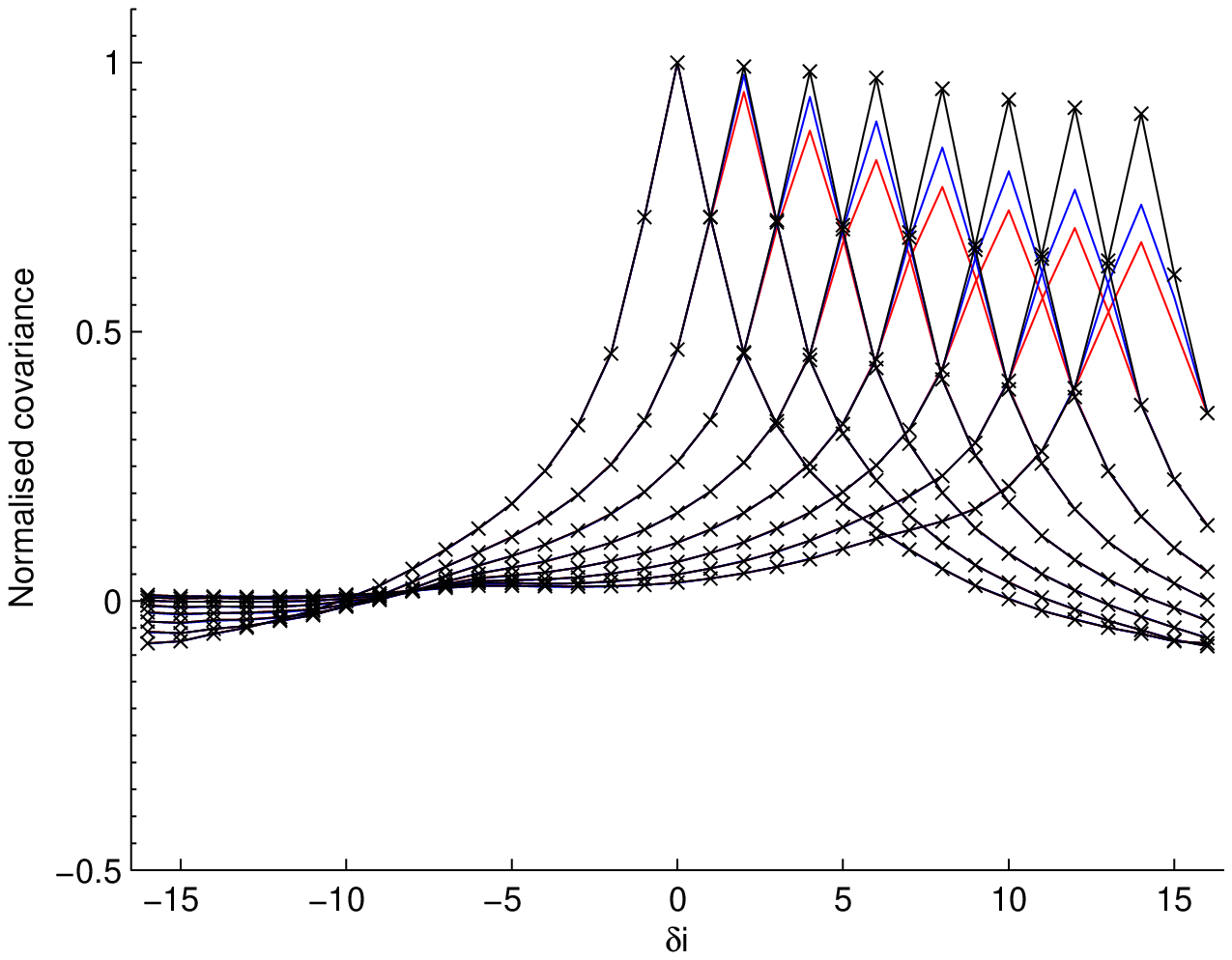}}}
      }
    \caption{Comparison of propagation effects on the normalized theoretical covariance response functions for increasing height ($H$) and different sub-aperture sizes, $w$. The covariance plots are (a) longitudinal and (b) transverse. The sub-aperture size $w=5.8~\rm{cm}$ with no propagation (black line with crosses); $w=11.6~\rm{cm}$ with propagation (blue line) and $w=5.8~\rm{cm}$ with propagation (red line). The plots show that propagation effects are lessened by the larger sized sub-aperture, $w=11.6cm$, but still significant $\sim30\%$. The theoretical covariance functions are plotted for even $\Delta$ with $H=\Delta \delta h$, where $\delta h = 1.29~\rm{km}$ and $H_{max}=20.6~\rm{km}$ ($\Delta=16$). The theoretical covariance functions for the SLODAR model are discrete valued, defined for integer values, $\delta i$, but plotted as continuous lines for clarity.   }
    \label{fig:SLODAR_Theoretical_Propagation}
  \end{center}
\end{figure}

In Fig.~\ref{fig:SLODAR_Theoretical_Propagation} we show the effects of propagation on the theoretical covariance impulse response functions for increasing height, $H$, and for sub-aperture sizes, $w=5.8~\rm{cm}$ and $w=11.6~\rm{cm}$. The covariance impulse functions are modelled for the ANU 17x17 SLODAR instrument using the methodology outlined by Butterley et al.~\cite{Butterley}, but with the modified power spectrum, Eq.~\ref{eqn:powerspec}. The $w=11.6~\rm{cm}$ size sub-apertures are modelled using a telescope pupil with twice the diameter ($D=2~\rm{m}$) compared to the $w=5.8~\rm{cm}$ size sub-apertures ($D=1~\rm{m}$), but impulse functions are identical as pupil geometry is unchanged~\cite{Butterley}. The propagation effects in decreasing the peak covariance values become more severe for increasing height,  $H=\Delta \delta h$, where in Fig.~\ref{fig:SLODAR_Theoretical_Propagation} have the values $\delta h = 1.29~\rm{km}$ and $H_{max}=20.6~\rm{km}$ ($\Delta=16$). The propagation effects are lessened for the larger sized sub-aperture, $w=11.6~\rm{cm}$, but are still significant.

\begin{figure}[htbp]
  \begin{center}
    \mbox{
      \subfigure[]{\scalebox{1.0}{\includegraphics[width=0.45\textwidth]{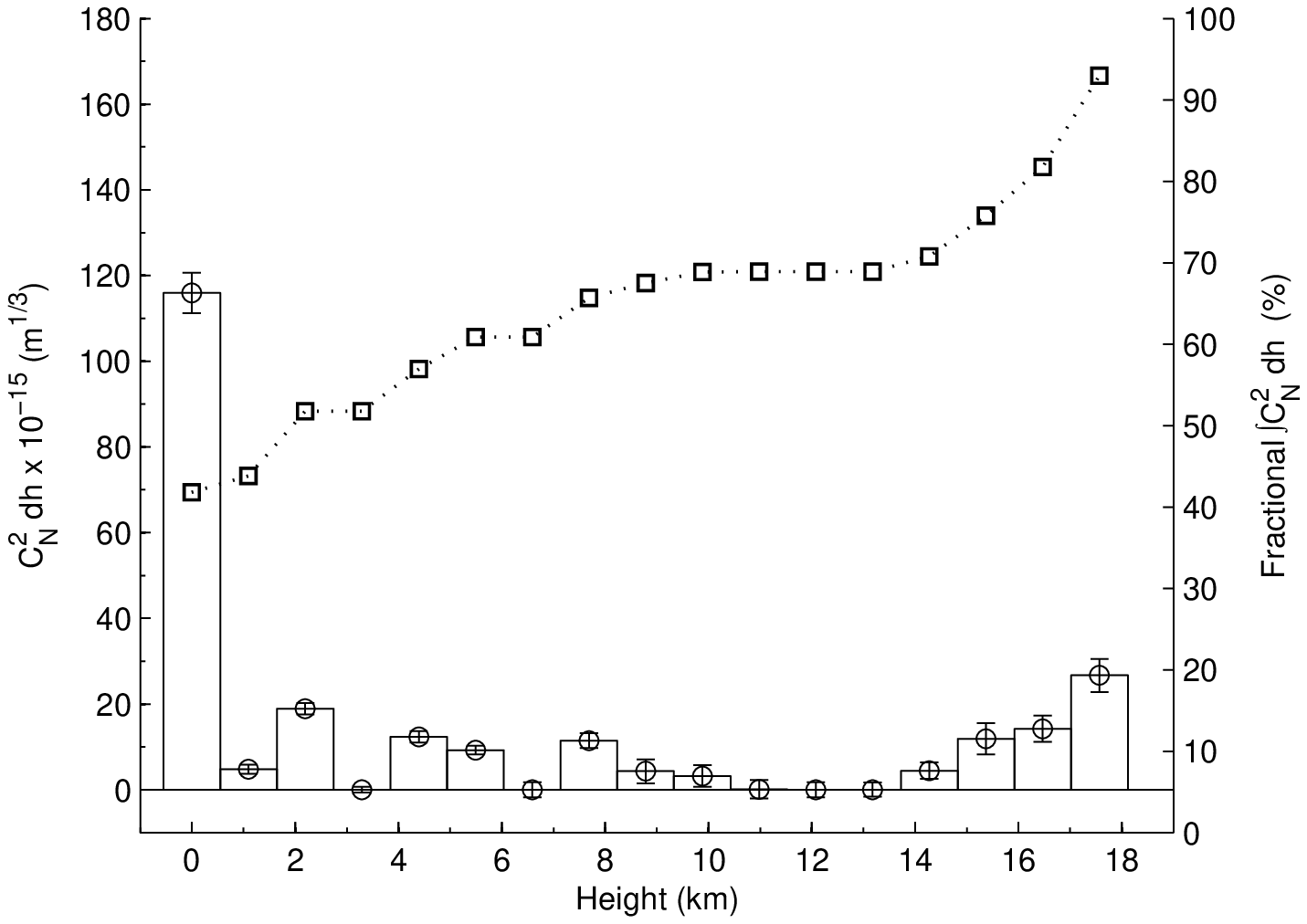}}} \quad
      \subfigure[]{\scalebox{1.0}{\includegraphics[width=0.45\textwidth]{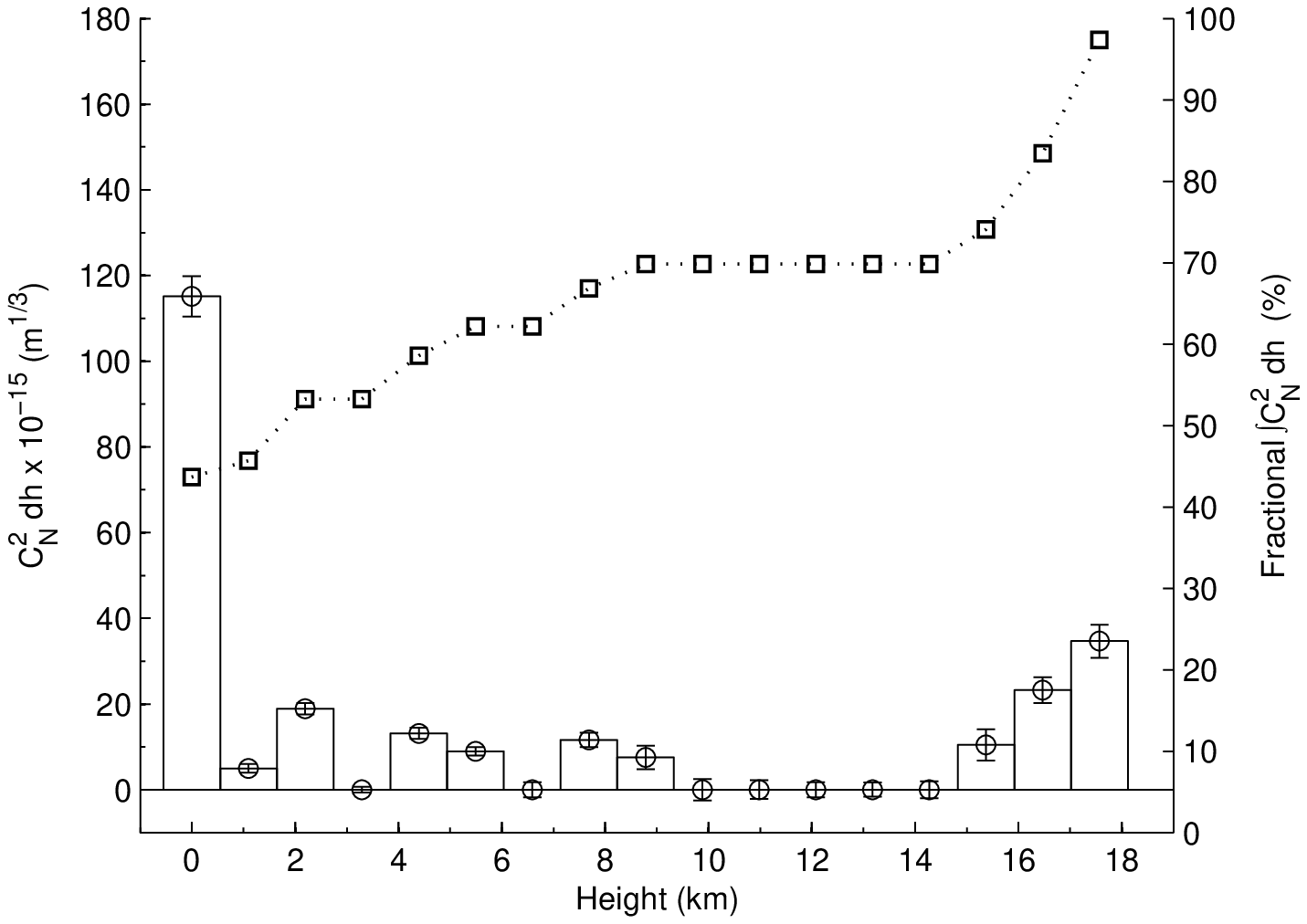}}}
      }
    \caption{A comparison between using the theoretical covariance impulse functions with (a) no propagation effects and (b) propagation effects on actual observational data taken 16:19 12 April 2006 (UTC) with ANU 17x17 SLODAR instrument on the ANU 40" telescope. The observational data was taken under excellent seeing, $r_{0}=18.2~\rm{cm}$ and exhibits significant high-altitude turbulence. The inclusion of propagation effects, (b), increases the strength of the highest turbulence, $H\sim16~\rm{km}$ by $\sim 25\%$ relative to (a), in agreement with Fig.~\ref{fig:SLODAR_Theoretical_Propagation}. The high-altitude turbulence $H>15~\rm{km}$ causes a steeper increase in the cumulative turbulence of (b) by $\sim 30\%$ compared to (a) $\sim 25\%$. The high-altitude of (b) appears more concentrated than (a). }
    \label{fig:SLODAR_Observation_Propagation}
  \end{center}
\end{figure}

In Fig.~\ref{fig:SLODAR_Observation_Propagation} we compare propagation effects on observational data taken 16:19 12 April 2006 (UTC) with the ANU 17x17 SLODAR instrument on the ANU 40" telescope. The 17x17 SLODAR instrument has relatively small sub-aperture size, $w = 5.8~\rm{cm}$, and therefore susceptible to propagation effects in the underestimation of high altitude turbulence. The observational data represents an individual run of the double star $\alpha~\rm{Cen}$, with angular separation, $\theta=9.44''$. The dataset consists of 2000 frames at 200~fps and exposure of 2~ms under excellent seeing, $r_{0}=18.2~\rm{cm}$, for Siding Spring. The dataset is selected as an example as it exhibits significant high-altitude turbulence, subsequently verified in the temporal-spatial cross-covariance data, moving at speeds $\sim 20~\rm{ms^{-1}}$. The centroid data was filtered with a 1~Hz high-pass FIR (Finite Impulse Response) filter to remove static mirror and dome seeing contributions from the ground-layer $\Delta=0$ measurement (see Section \ref{sec:mirrorturb}).The estimation of the $C_{N}^2(h)dh$ profile was implemented by fitting of the transverse (T) theoretical covariance functions for a Kolmogorov turbulence power spectrum, $\beta=11/3$. From Fig.~\ref{fig:SLODAR_Observation_Propagation} the inclusion of propagation effects increase the strength of the high-altitude turbulence $H>15~\rm{km}$ by $\sim 25\%$.

\section{Removal of Mirror and Dome Turbulence}

\label{sec:mirrorturb}
\begin{figure}[htbp]
  \begin{center}
    \mbox{
      \subfigure[]{\scalebox{1.0}{\includegraphics[width=0.45\textwidth]{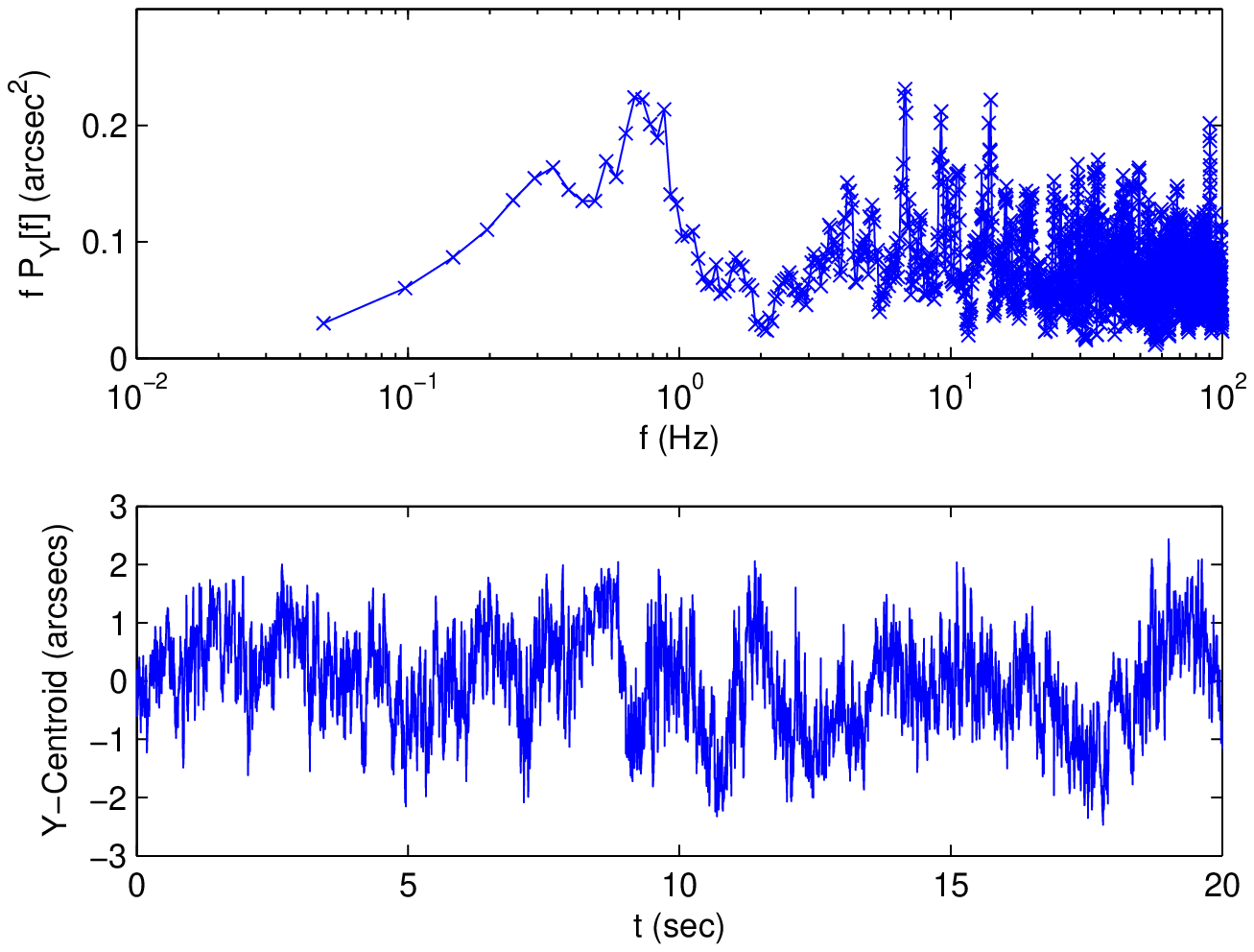}}} \quad
      \subfigure[]{\scalebox{1.0}{\includegraphics[width=0.45\textwidth]{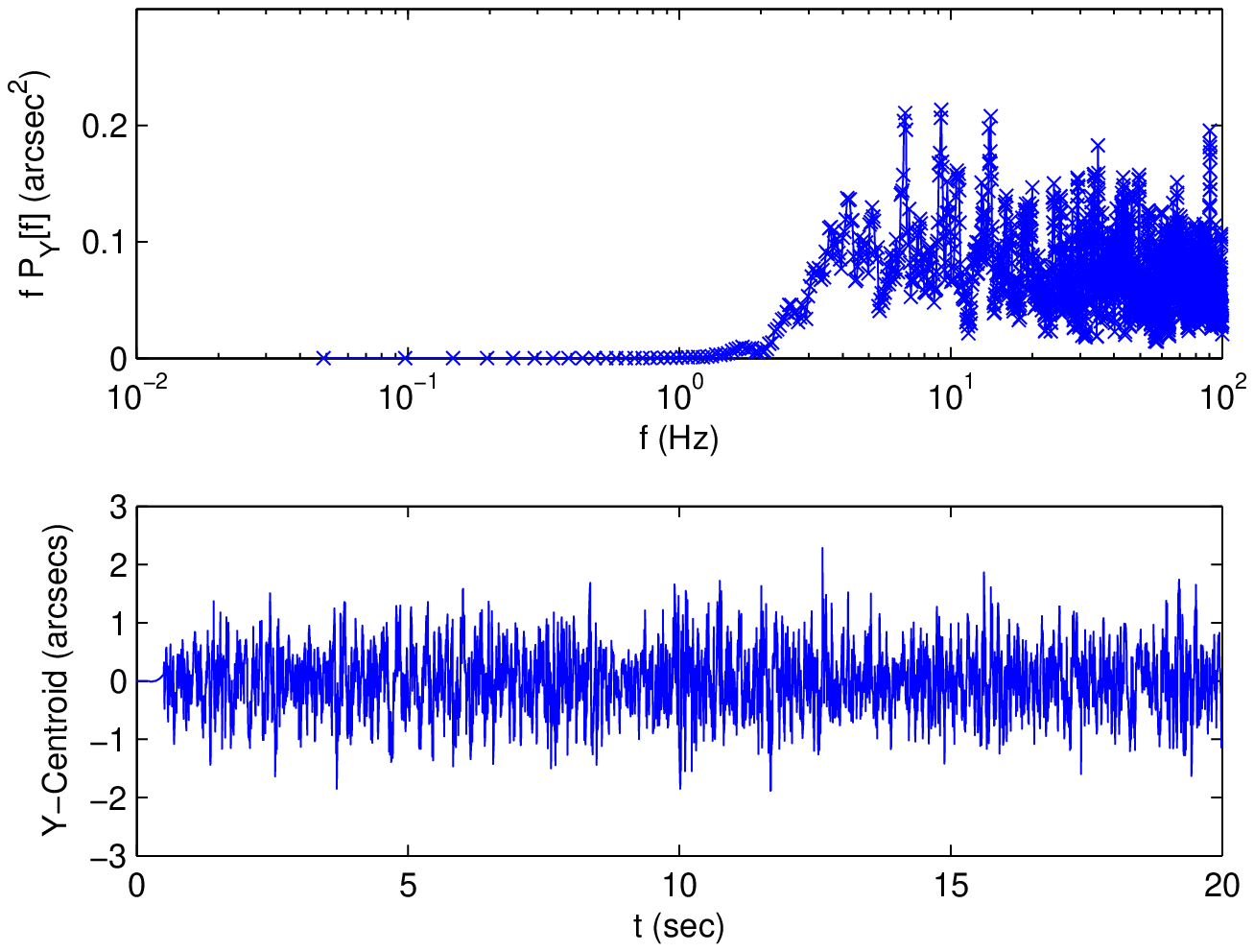}}}
      }
    \mbox{
      \subfigure[]{\scalebox{1.0}{\includegraphics[width=0.45\textwidth]{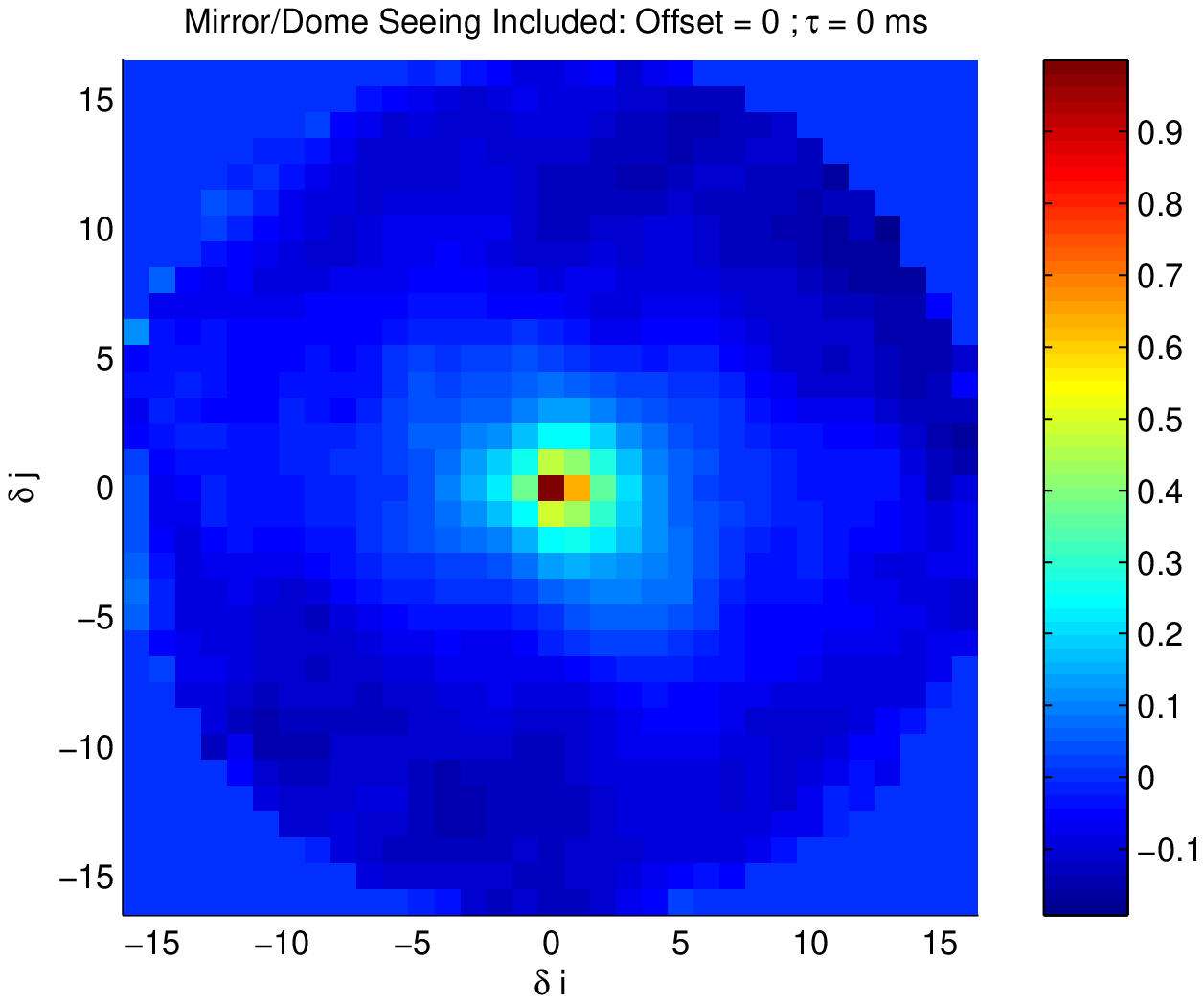}}} \quad
      \subfigure[]{\scalebox{1.0}{\includegraphics[width=0.45\textwidth]{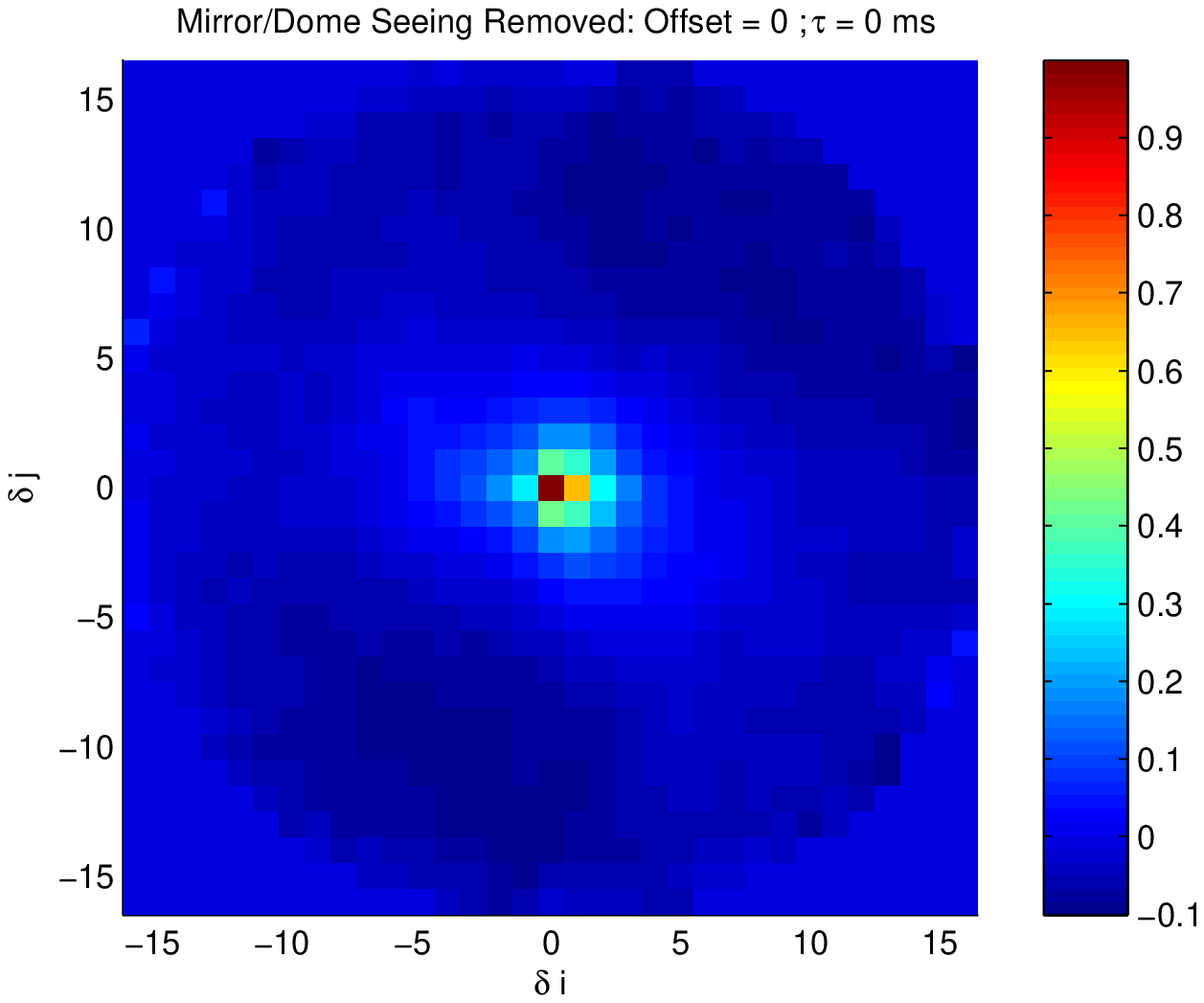}}}
      }
    \caption{Plots (a) and (c) are observational data containing significant amounts of mirror and dome seeing that cause over-estimating the contribution of the atmospheric ground-layer. Plots (b) and (d) are observational data with mirror and dome seeing removed by application of a high-pass filter with cut-off of 2~Hz to the centroid data streams. Plots (a) and (b) are the Spectral Energy Density of the Y-centroid slope data (top) and Y-centroid data stream (bottom) for star A sub-aperture index [$i=2$, $j=5$]. Plots (c) and (d) are AVI animations (size: (c) 579 KB and (d) 581 KB) of the spatial-temporal 2-D cross-covariance sequences for offset lags of 0 to 20 frames, each offset lag is 5~ms apart. The cross-covariance sequences are for X-centroid and Y-centroid data, and normalized to the zero spatial offset peak, [$\delta i=0$, $\delta j=0$], for zero offset lag, $\tau=0~\rm{ms}$. The contribution of mirror and dome seeing to the ground-layer measurement ([$\delta i=0$, $\delta j=0$], $\tau=0~\rm{ms}$) of plot (c) is about 48\%, observed as excess residual for [$\delta i=0$, $\delta j=0$] for $\tau=50~\rm{ms}$. Plots (a)-(d) reference the observational dataset of $\alpha~\rm{Cen}$ consisting of 20~s of data at 200~fps (4000 frames), taken 12:43 20 June 2006 UTC with the ANU 17x17 SLODAR instrument on the ANU 40" telescope.}
    \label{fig:SLODAR_Mirror_Turbulence}
  \end{center}
\end{figure}

Mirror and dome turbulence manifests as a distinct separate component in the power spectrum of angular tilt. Mirror and dome turbulence appears at low temporal frequencies. The mirror and dome turbulence contaminates the ground-layer measurement of the turbulence profile. In some cases the mirror and dome turbulence can be falsely over-estimate the contribution of the ground-layer relative to the free-atmosphere layers. The effects of mirror and dome seeing on observational data are shown in Fig.~\ref{fig:SLODAR_Mirror_Turbulence}. The observational data is taken with the ANU 17x17 SLODAR instrument on the ANU 40" telescope.

The mirror and dome turbulence shows strong correlation in the zero spatial offset of the temporal spatial cross-correlation frame data for time lags over 100~ms. Following a suggestion by R.~W. Wilson (private communication with C.~Jenkins~\cite{Wilsona}) we have found that the mirror and dome turbulence can be removed from observational data by applying a high pass filter to centroid data with cut-off approximately in the range of 1-2~Hz.

\section{Improving the Vertical Resolution}
\label{sec:verticalres}
The nominal height resolution of SLODAR is $\delta h=w/\theta$, where $w$ is the sub-aperture width and $\theta$ is the angular separation of the observed double star. To improve the height resolution, $\delta h$, assuming fixed $\theta$ and fixed exposure time, $\tau$, is to reduce the size of the sub-apertures, $w$. The number of signal photons per $\tau$ is directly proportional to $w^2$ so reducing $w$ will cause photon starvation and hence restrict observations to the brighter double stars (few in number). The minimum useable sizes of $w$ range from $w=5~\rm{cm}$ for the portable ESO 8x8 SLODAR system~\cite{Wilson} and $w=5.8~\rm{cm}$ for our ANU 17x17 SLODAR system. Reducing $w$ will also present a number of second order effects, including less sensitivity to high altitude layers due to typical high wind speeds and propagation effects. The high wind speeds, $v$ will reduce the variance of the observed tilts contributed by increasing the effective sampling distance of wavefront tilts from $w$ to $\tau v$ when $\tau v > w$~\cite{Martin}. Propagation effects (see Section \ref{sec:propagation}) will reduce or null power in the phase fluctuation power spectrum at spatial frequencies near the Fresnel length, or near $w$, resulting in reduction in the variance of the observed tilts contributed by the layer. Hence a physical limit exists for minimum size of $w$ and therefore a minimum height resolution, $\delta h$, to ensure a satisfactory performance of SLODAR.

Therefore, we propose the concept of Generalized SLODAR that improves the height resolution of SLODAR by combining measurements taken at regularly spaced SHWFS conjugation heights at the nominal resolution, $\delta h$. The SHWFS conjugation heights are a fractional amount of the nominal height resolution, $\delta h$. By combining $N_G$ datasets at regularly spaced SHWFS conjugation heights, new information is provided about the atmospheric turbulence, and  is possible to achieve a Generalized SLODAR height resolution, $\delta h^*= \delta h/ N_G$.

We begin this section by describing the process of retrieving an estimated profile of the atmospheric turbulence from observational data. We then introduce the notation and summarize the results of theoretical covariance impulse functions derived in the paper by Butterley et al.~\cite{Butterley}. We then extend their results to the case of Generalized SLODAR by defining a new set of coordinates and the methodology for combining measurements and retrieval of the super-resolution turbulence profile. We validate the methodology by showing the results of a numerical simulation of resolving two phase screens closely separated in height. We then apply the methodology to observational data and demonstrate an improvement by a factor of two in height resolution.

The SLODAR technique does not measure the atmospheric turbulence profile directly but needs to be recovered from the wavefront slope cross-covariance data from stars A and B. The atmospheric turbulence profile is an internal property of the system that can be estimated by fitting modelled theoretical covariance impulse response functions of thin turbulent layers spaced by $\delta h$. This model assigns equal a priori probability to all heights (unbiased model). The fitting procedure can be modelled as system of linear equations in matrix form, $\mathbf{Ax} = \mathbf{b}$. $\mathbf{A}$ is the kernel matrix with column vectors corresponding the theoretical covariance impulse response functions. $\mathbf{b}$ is an ensemble average of the observed atmospheric turbulence covariance profile (that also includes systematic and statistical noise), represented as a column vector. $\mathbf{x}$ is the quantity that we seek, an estimate of the atmospheric turbulence profile, represented as a column vector of strengths of each thin layer.

We must note that the estimate of the atmospheric turbulence, $\mathbf{x}$, is based on the input data, $\mathbf{b}$, assumptions made by the model, $\mathbf{A}$, and the process to recover $\mathbf{x}$, (inversion models). The system is over-determined as there are more equations than variables so matrix $\mathbf{A}$ cannot be directly inverted and a least squares solution is sought. The system solution, $\mathbf{x}$, can be found by least squares inversion,$\mathbf{x} = \mathbf{A^{+}b}$, where $\mathbf{A^+}$ is the pseudo inverse of $\mathbf{A}$. However, we note as $\mathbf{b}$ contains unwanted noise so the system may be un-stable and hence the solution, $\mathbf{x}$, invalid.

We can improve the inversion model by using the prior information that the layer strengths are a positive quantity, $\mathbf{x>0}$ and that $\mathbf{b}$ is possibly corrupted with Gaussian noise. Such an inversion model is the Non-Negative Least Squares (NNLS) algorithm. We have found through simulation that the NNLS algorithm recovers the input atmosphere model more accurately than other regularization algorithms, such as MAXENT and Tikhonov regularization. The NNLS algorithm is implemented in MATLAB as the routine \emph{lsqnonneg} and performs well on compact sources (minimal smoothing). Hence the NNLS algorithm suitable with the thin-layer model assumption of the atmosphere, as verified with typical high resolution measurements of atmospheric turbulence~\cite{Azouit}.

For SLODAR, the 2-D theoretical covariance impulse response function to a turbulent layer at altitude H for the wavefront tilts in longitudinal (L) direction, given that the double star separation axis is aligned with SHWFS x-axis, has been derived by Butterley et al.~\cite{Butterley}:

\begin{equation}
X_L (\Delta ,\delta i,\delta j) = \frac{1}
{{N_{cross} }}\sum\limits_{valid\text{ }i,j,i^{'} ,j^{'} } {C^{'x} _{i,j,i^{'} ,j^{'} } (\Delta )}
\end{equation}

The function $C^{'x} _{i,j,i^{'} ,j^{'} } (\Delta )$  describes the theoretical covariance of x-directional slopes for a cross-pair of sub-apertures with lateral pupil spatial offset ($\delta i$,$\delta j$) and layer height, $H= \Delta \delta h$, after global tilt subtraction. The number of cross-pair lenslets having the same lateral pupil spatial offset ($\delta i$,$\delta j$) for a given layer height, $H$,is denoted by $N_{cross}$. The notation used to describe the theoretical covariance impulse response function, $X_L (\Delta ,\delta i,\delta j)$, is shown in Fig.~\ref{fig:SLODAR_Geometry}. The indices [$i$,$j$] refer to the lenslet index for star A and [$i'$,$j'$] for star B. A cross-pair of lenslets has a lateral pupil spatial offset defined by ($\delta i$,$\delta j$)=($i'-i$,$j'-j$), specified  in units of the sub-aperture width, $w$. The lenslet index, $i$, takes on integer values $i = \{1, 2,..., N \}$, where $N$ is the number of lenslets mapped across the diameter of the telescope pupil. Likewise for indices $j$, $i'$ and $j'$. The lateral pupil spatial offset, $\delta i$, takes on integer values $\delta i = \{ 1-N, 2-N, ... , 0 , 1, ... N-2, N-1 \}$, specified in units of sub-aperture width, $w$. Likewise for $\delta j$.

If the double star separation axis is aligned with the SHWFS x-axis, then processing is simplified by considering the covariance function of the tilts in longitudinal (L) or transverse (T) directions relative to the lateral pupil spatial offset,  ($\delta i$,$\delta j$),  (units of $w$). A turbulent layer height at $H$ corresponds to a lateral pupil spatial separation, $\Delta = H \theta / w$ of telescope pupils, specified in units of the sub-aperture width, $w$. The lateral pupil spatial separation, $\Delta$, is an offset of the projected telescope pupils along the x-direction at the layer altitude, $H$, and takes on integer values $\Delta = \{ 0,1,2,...,N-1 \}$. The physical separation of a pair of sub-apertures with a spatial offset ($\delta i$,$\delta j$)  projected on a layer at height, $H$, is then ($u_x$, $u_y$) where $u_x=|\delta i+\Delta|w$ and $u_y=|\delta j|w$, is used by $C^{'x} _{i,j,i^{'} ,j^{'} } (\Delta )$ function.  Hence the estimated strengths of the layers are defined with height bins of widths $\delta h$ and centered at $\Delta \delta h$.  However, the practical height resolution, $\delta h$, may be poorer depending on the signal-to-noise ratio of observational data and the inversion model implemented to recover the estimated strengths.

The 2-D theoretical covariance impulse response function, $X_L (\Delta ,\delta i,\delta j)$, is an accurate model, and takes into consideration the pupil geometry (mapping of circular or square sub-apertures on the annular telescope), turbulence power spectrum ($\beta$, $L_o$) and effects of 'global' tilt subtraction (tilt anisoplanatism) required to remove telescope tracking errors. The parameters $\Delta$ and ($\delta i$,$\delta j$) are integer valued and hence $X_L (\Delta ,\delta i,\delta j)$ is a discrete function that models the impulse response of equally spaced thin layers with height, $\Delta \delta h$. The discrete impulse response function $X_L (\Delta ,\delta i,\delta j)$ is in a format that is compatible with the discrete observational covariance profile $\mathbf{C}_{L,k}^{'x,obs} (\delta i,\delta j)$. Hence the discrete function $X_L (\Delta ,\delta i,\delta j)$ can be specified in matrix form, $\mathbf{A}$, to model the system as a set of linear equations, $\mathbf{Ax} = \mathbf{b}$, and then inverted to solve for layer strengths, $\mathbf{x} = \mathbf{A^{+}b}$. To further explain the process the discrete observational covariance profile $\mathbf{C}_{L,k}^{'x,obs} (\delta i,\delta j)$ can be modelled as a linear equation in the form

\begin{equation}
\mathbf{C}_{L,k}^{'x,obs} (\delta i,\delta j) = x_0 X_L (0,\delta i,\delta j) + x_1 X_L (1,\delta i,\delta j) +  \cdots  + x_{N - 1} X_L (N - 1,\delta i,\delta j)
\end{equation}

where, as noted, the combination of $\delta i$ and $\delta j$ maps the theoretical covariance impulse response, $X_L (\Delta ,\delta i,\delta j)$, of a particular height, $H=\Delta w / \theta$. Expressing as a set of linear equations, $\mathbf{Ax} = \mathbf{b}$, where $\mathbf{x}$ is a column vector of layer strengths:

\begin{eqnarray}
\left[ {\begin{array}{*{20}c}
   \begin{gathered}
  col\left\{ {X_L (0,\delta i,\delta j)} \right\} \hfill \\
   \hfill \\
   \hfill \\
   \hfill \\
\end{gathered}  & \begin{gathered}
  col\left\{ {X_L (1,\delta i,\delta j)} \right\} \hfill \\
   \hfill \\
   \hfill \\
   \hfill \\
\end{gathered}  & \begin{gathered}
  ... \hfill \\
   \hfill \\
   \hfill \\
   \hfill \\
\end{gathered}  & \begin{gathered}
  col\left\{ {X_L (N - 1,\delta i,\delta j)} \right\} \hfill \\
   \hfill \\
   \hfill \\
   \hfill \\
\end{gathered}   \\

 \end{array} } \right]\left[ \begin{gathered}
  x_0  \hfill \\
  x_1  \hfill \\
  . \hfill \\
  x_{N - 1}  \hfill \\
\end{gathered}  \right]
\nonumber \\
= \left[ \begin{gathered}
  col\left\{ {\mathbf{C}_{L,k}^{'x,obs} (\delta i,\delta j)} \right\} \hfill \\
   \hfill \\
   \hfill \\
   \hfill \\
\end{gathered}  \right]
\end{eqnarray}

where $col\{\mathbf{x}\}$ denotes the process that serializes the 2-D data, $\mathbf{x}$, into a 1-D column vector by stacking columns of $\mathbf{x}$ with increasing $\delta i$.

The covariance impulse function is further simplified by taking a 1-D cut along $y=0$ of the 2-D theoretical covariance function, $X_L (\Delta ,\delta i,\delta j)$, or by setting $j'=j$ or $\delta j=0$.

\begin{equation}
X_L (\Delta ,\delta i) = \frac{1}
{{N_{cross} }}\sum\limits_{valid\text{ }i,j,i^{'} } {C^{'x} _{i,j,i^{'} ,j} (\Delta )}
\end{equation}

The 1-D theoretical covariance function,$X_L (\Delta ,\delta i)$ , is calculated for integer valued lateral pupil spatial separations, $\Delta = \{ 0,1,2,...,N-1 \}$ and the condition  $\Delta = 0$  corresponds to completely overlapped telescope pupils projected on the SHWFS. This configuration is when the SHWFS is conjugated to the telescope pupil ($h_0$=0~km), refer Fig.~\ref{fig:SLODAR_Geometry}.

Information between the nominal height bins, $\Delta \delta h$, can be found at non-integer lateral pupil spatial separations, $\Omega_k = \Delta + \eta_k= \{ 0+\eta_k,1+\eta_k,2+\eta_k,...,N-1+\eta_k \}$, where $\eta_k$ takes values between 0 and 1, where $k$ is the index of the group of  $N_G$ Generalized SLODAR datasets, $k = \{0,1,...,N_G-1\}$. The value $\Omega_k$  can be obtained by moving the SHWFS conjugation height, $h_0$, upwards by fractional amounts of the height resolution, $h_0^*= \eta_k\delta h$, and is illustrated in Fig.~\ref{fig:SLODAR_Geometry}. Moving the conjugation height, $h_0^*$, results in a lateral pupil spatial offsets, $\delta m_k = \delta i + \eta_k$ and $\delta m_k = \eta_k$ for $\delta i =0$, corresponding to a lateral pupil spatial separations, $\Omega_k = \eta_k$. Hence telescope pupils are no longer completely overlapped at the SHWFS but separated by a fractional amount of a lenslet. The non-integer lateral pupil spatial separations, $\Omega_k$, can be thought of sampling new and unique spatial offsets, $\delta i + \eta_k$, in the telescope pupil.

\begin{figure}[htbp]
\centering
  \includegraphics[width=0.95\textwidth]{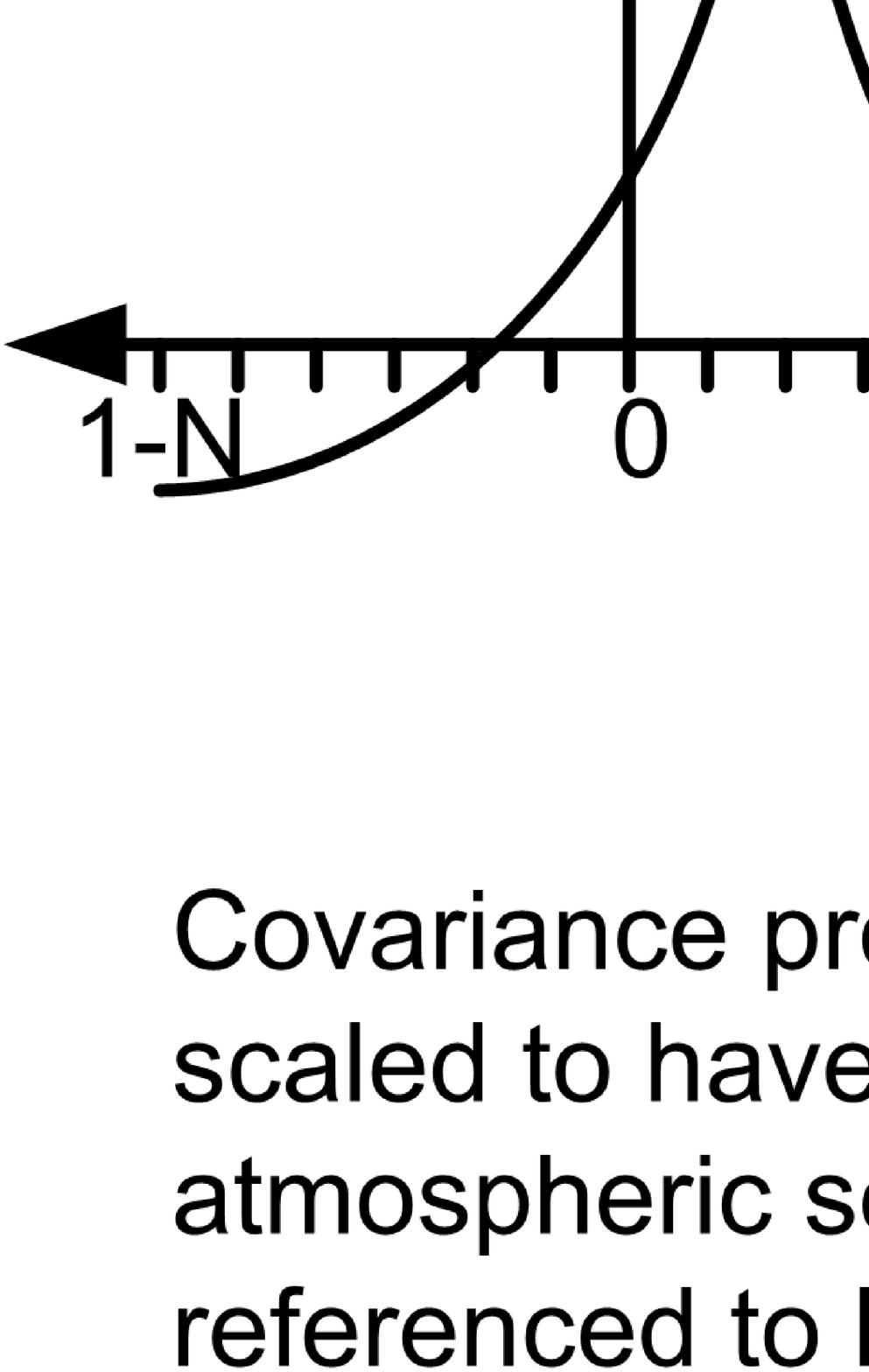}
  \caption{The data reduction methodology for Generalized SLODAR}\label{fig:SLODAR_Generalized_Methodology}
\end{figure}

The aim of Generalized SLODAR is to reconstruct a super-resolution turbulence profile by combining several datasets, $k$, having unique lateral pupil spatial separations, $\Omega_k$, and with equal height resolutions, $\delta h$. The methodology for Generalized SLODAR is shown in Fig.~\ref{fig:SLODAR_Generalized_Methodology}.

To provide an unbiased super-resolution profile the lateral pupil spatial separations, $\Omega_k$, must be equally spaced and hence require $\eta_k$ to also be equally spaced. The fractional spacings, $\eta_k$ are then given by $\eta_k=k/N_G$  and therefore $\eta_k = (1/ N_G ) \{0,1,2,..., N_G-1\}$.

We denote the observed global-tilt removed covariance profile in the longitudinal (L) direction of double star having a separation axis aligned along the x-axis of the SHWFS, for an individual Generalized SLODAR dataset, $k$, to be $\mathbf{C}_{L,k}^{'x,obs} (\delta i)$. Note that the symbol defined for the observed covariance profile $\mathbf{C}_{L,k}^{'x,obs} (\delta i)$ should be clearly distinguished from the theoretical covariance function for a cross-pair of lenslets, $C^{'x} _{i,j,i^{'} ,j^{'} } (\Delta )$. We now need to transform the observed covariance profile from a local lenslet-based coordinate system to a global coordinate system,  $\mathbf{C}_{L,k}^{'x,obs} (\delta m_k)$,  referenced to $\eta_k=0$, or lateral spatial offsets in the telescope pupil at $h_0=0$. The global coordinate system of an individual Generalized SLODAR dataset, $k$, is defined as $\delta m_k = \delta i + \eta_k$, specified  in units of the sub-aperture width, $w$. To construct the observed super-resolution covariance profile,  $\mathbf{C}_{L}^{*'x,obs} (\delta m^*)$, requires the $\mathbf{C}_{L,k}^{'x,obs} (\delta m_k)$ profiles to be first scaled to normalize fluctuations in seeing and then interleaved. The scaling parameter, $a_k$, normalizes $\mathbf{C}_{L,k}^{'x,obs} (\delta m_k)$ to have equal seeing and hence remove any bias effects, and defined as

\begin{equation}
a_k  = \frac{{\mathbf{A}_{L,0}^{'x,obs} (\delta i = 0)}}
{{\mathbf{A}_{L,k}^{'x,obs} (\delta i = 0)}}
\end{equation}

where ${\mathbf{A}_{L,k}^{'x,obs} (\delta i = 0)}$  refers to the peak of the centroid-noise removed auto-covariance function for dataset $k$, and proportional to the total atmospheric seeing. For most cases the scaling parameter, $a_k$, is close to unity,  $a_k\approx1$.  The observed super-resolution covariance profile, $\mathbf{C}_{L}^{*'x,obs} (\delta m^*)$, is then

\begin{equation}
\mathbf{C}_L^{*'x,obs} (\delta m^* ) = \bigcup\limits_k {a_k \mathbf{C}_{L,k}^{'x,obs} (\delta m_k )}
\end{equation}

where

\begin{equation}
\delta m^*  = \bigcup\limits_k {\delta m_k }
\end{equation}

is the combined spatial offsets relative to the Classical SLODAR, expanded form:

\begin{equation}
\delta m^*  = \left\{ {\delta m_0  \cup \delta m_1  \cup ... \cup \delta m_{N_G  - 1} } \right\}
\end{equation}

The pupil spatial separations are denoted by:

\begin{equation}
\Omega _k  = \Delta  + \eta _k
\end{equation}

where

\begin{equation}
\Omega ^*  = \bigcup\limits_k {\Omega _k }
\end{equation}

is the combined pupil spatial separations relative to the Classical SLODAR, expanded form:

\begin{equation}
\Omega ^*  = \left\{ {\Omega _0  \cup \Omega _1  \cup ... \cup \Omega _{N_G  - 1} } \right\}
\end{equation}

we now need to calculate the super-resolution theoretical function:

\begin{equation}
X_L^* (\Omega ^* ,\delta m^* ) = \frac{1}
{{N_{cross} }}\sum\limits_{valid\text{ }m,l,m^{'} } {C^{'x} _{m,l,m^{'} ,l} (\Omega ^* )}
\end{equation}

where $m$ and $l$ are now indices that reference a higher sampled SHWFS at fractional spacings of a sub-aperture, $w^*=w/N_G$ with total samples, of $N^*= N_G N$. Due to the complexity and time required to compute $X_L^* (\Omega ^* ,\delta m^* )$ it is best to approximate with interpolation methods. Through numerical simulations involving phase screens, it found that cubic interpolation method is suitable for the $X_L^* (\Omega ^* ,\delta m^* )$ function and spline interpolation for $X_T^* (\Omega ^* ,\delta m^* )$ function.

The theoretical covariance function,  $X_L^* (\Omega ^* ,\delta m^* )$, can now be constructed in matrix form, $\mathbf{A}$, to model the system as a set of linear equations, $\mathbf{Ax} = \mathbf{b}$, and then inverted to solve for layer strengths,  $\mathbf{x} = \mathbf{A^{+}b}$. Due to the larger size of the matrix $\mathbf{A}$, it best to use a positively constrained, $\mathbf{x>0}$ inversion method for compact sources (minimal smoothing to $\mathbf{x}$), such as the Non-Negative Least Squares (NNLS) algorithm implemented as the MATLAB iterative routine \emph{lsqnonneg}.

From theoretical and numerical simulations the technique is successful for $N_G=3$. For $N_G=6$ the results are progressively poorer due to a larger matrix being increasingly sensitive to noise.

We validate the Generalized SLODAR methodology presented in this paper by showing the results of a numerical simulation for $N_G=3$ resulting in an effective height resolution, $\delta h^*= \delta h/ 3$. We confirm this by clearly separating two phase screens separated in height by $h=2 \delta h^*$.

\begin{table}
  \begin{center}
\begin{tabular}{|l|l|l|l}
\cline{1-3}
\textbf{Parameter} & \textbf{Value} & \textbf{Description} &  \\
\cline{1-3}
$\theta$ & 9.44" & double angular star separation &  \\
$\lambda$ & 0.5e-6 & mean wavelength &  \\
$D$ & 1.02 m & telescope diameter &  \\
$O$ & 0.45 & secondary / primary obstruction ratio &  \\
$w$ & 0.06 m & sub-aperture width (square) &  \\
$N$ & 17 & number of sub-apertures across telescope diameter &  \\
$\delta x$ & 1 cm/pixel & waverfront and pupil sampling &  \\
$\delta x_h$ & 200 m/pixel & vertical resolution of pupil sampling &  \\
$N_{frames}$ & 4000 & number of independent atmospheric realizations &  \\
$Propagation$ & Geometrical & phase screens added together for pupil wavefront &  \\
$H_1$ & 6800 m & height of phase screen for layer 1 &  \\
$H_2$ & 7600 m & height of phase screen for layer 2 &  \\
$r_0$ & 0.3 m & total integrated seeing &  \\
$\beta$& 11/3 & power law of phase power spectrum (Kolmogorov) &  \\
$N_G$ & 3 & number of generalized datasets &  \\
$\eta_k$ & { 0, 1/3, 2/3 } & fractional pupil offsets for generalized datasets &  \\
$\delta h$ & 1200 m & nominal resolution of each generalized dataset &  \\
$\delta h^*$ & 400 m & super-resolution of combined  profiles &  \\
\cline{1-3}
\end{tabular}
\caption{\label{tab:numparams}Parameters for the numerical simulation}
  \end{center}
\end {table}

We model the simulation after the double star $\alpha~\rm{Cen}$ and the ANU 17x17 SLODAR instrument on the ANU 40" telescope. The parameters of the simulation are listed in Tab.~\ref{tab:numparams}. The Generalized SLODAR is simulated by sequentially moving the layers down in vertical height by $\delta h^*$ or 400m for each fractional generalized pupil offsets, $\eta_k$. A lateral  pixel offset of 1cm corresponds to vertical height of 200m. Therefore, for each dataset, decreasing the separation of telescope pupils of star A and star B as projected onto the phase screens $H_1$ and $H_2$ by two pixels (2x200m) achieved Generalized SLODAR. The wavefronts for each star at the SHWFS is calculated by extracting the part of the phase screen that the pupils project on and then adding together for each layer $H_1$ and $H_2$.

\begin{figure}[htbp]
  \begin{center}
    \mbox{
      \subfigure[]{\scalebox{1.0}{\includegraphics[width=0.45\textwidth]{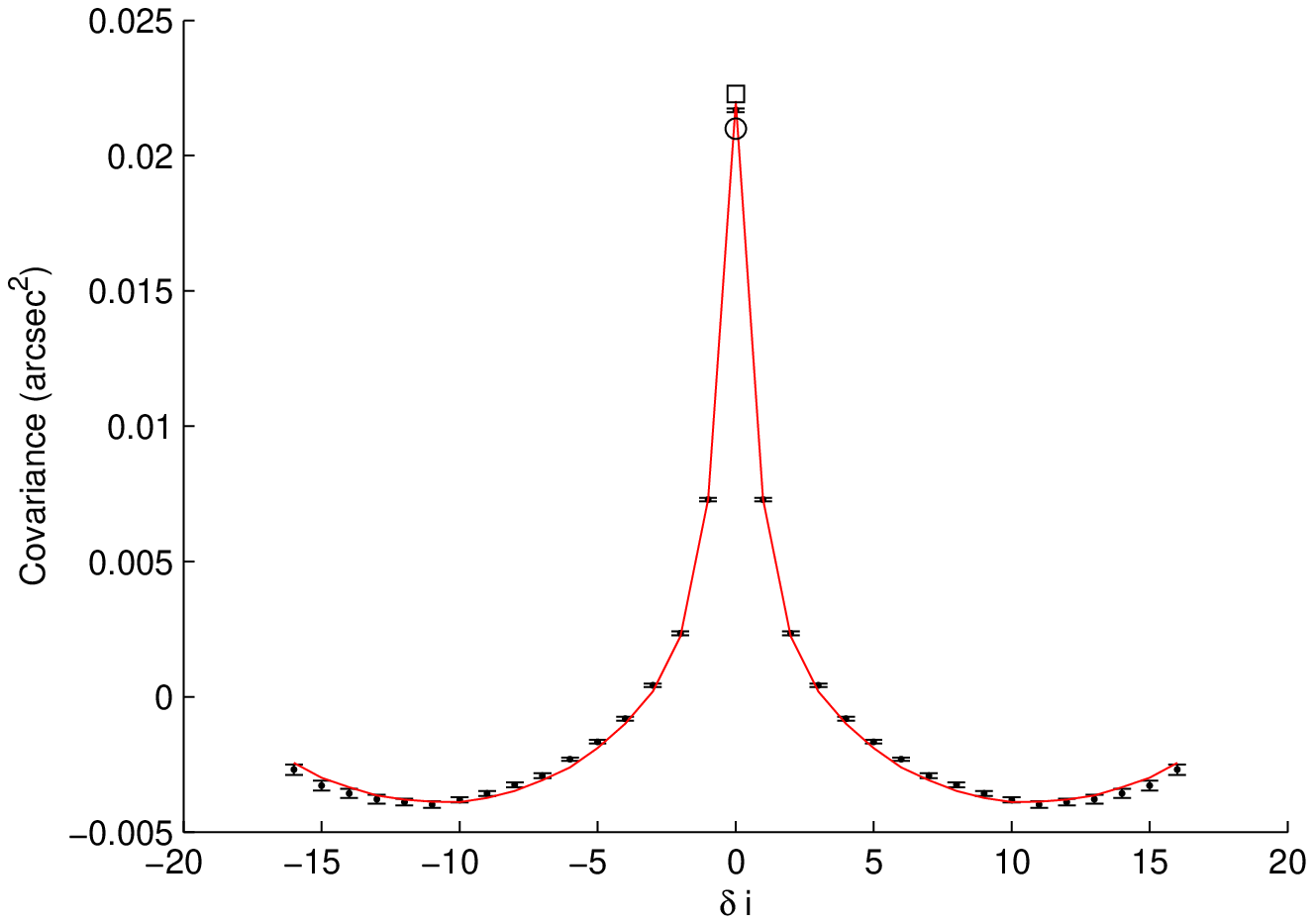}}} \quad
      \subfigure[]{\scalebox{1.0}{\includegraphics[width=0.45\textwidth]{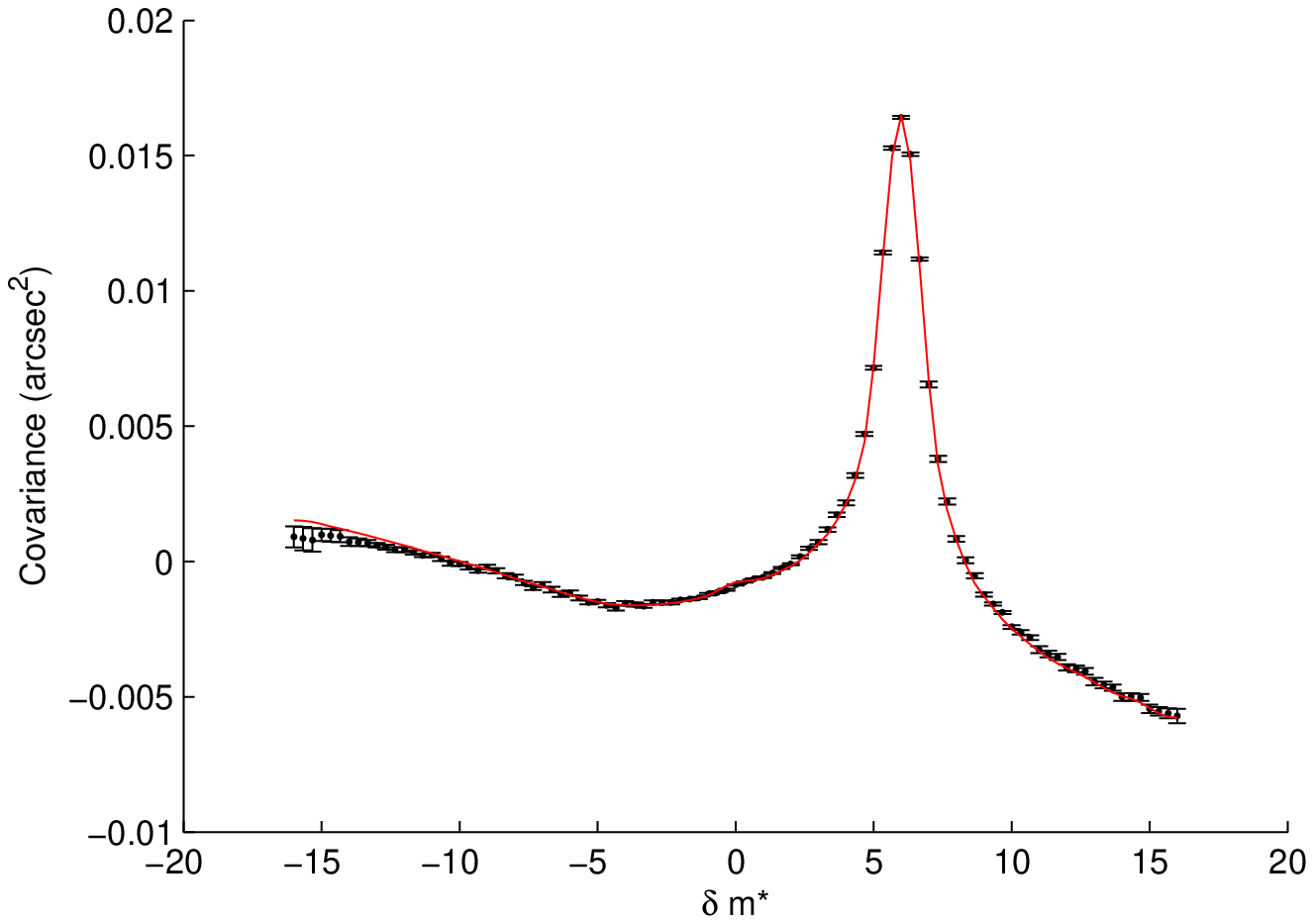}}}
      }
    \mbox{
      \subfigure[]{\scalebox{1.0}{\includegraphics[width=0.45\textwidth]{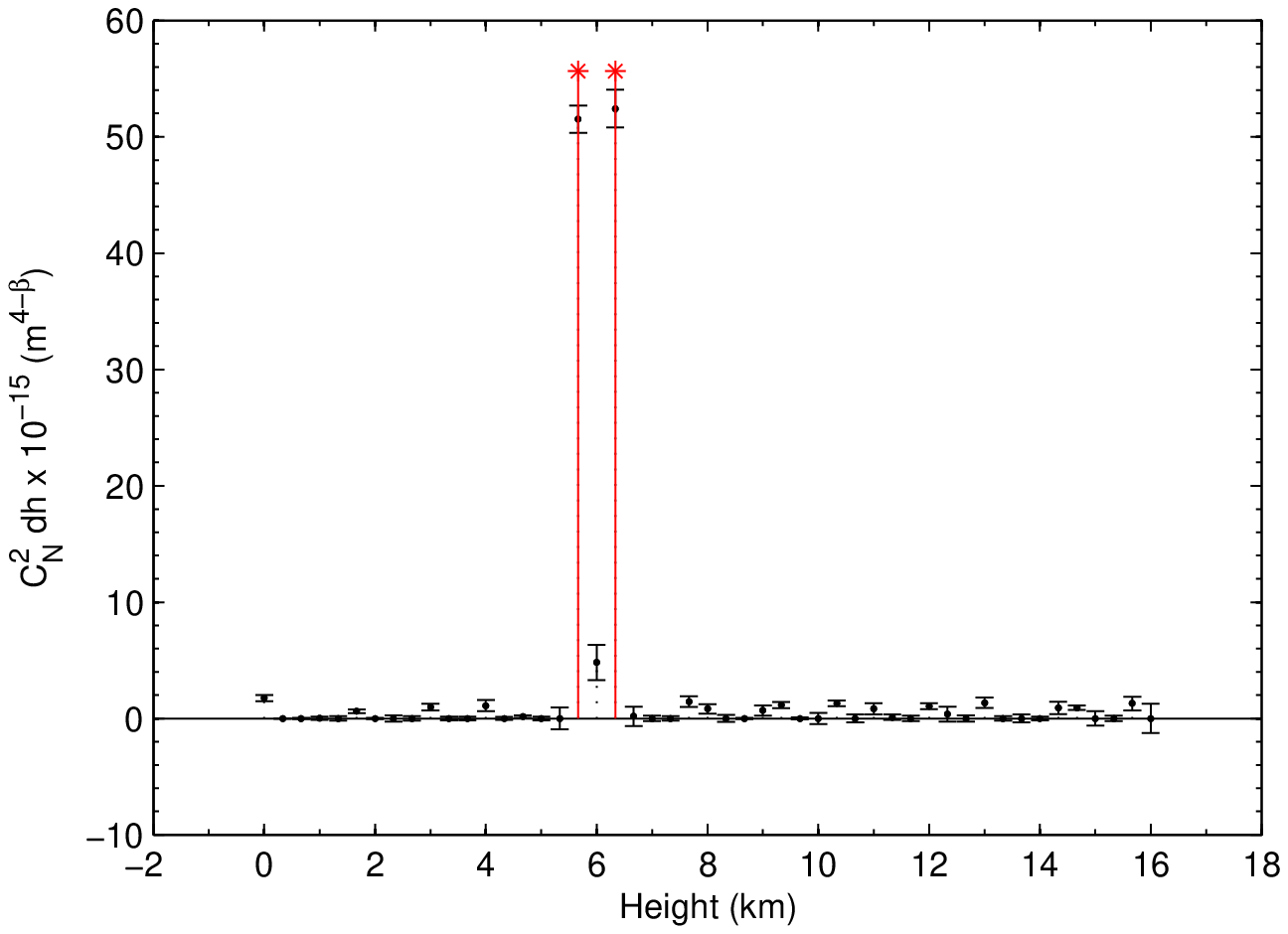}}} \quad
      \subfigure[]{\scalebox{1.0}{\includegraphics[width=0.45\textwidth]{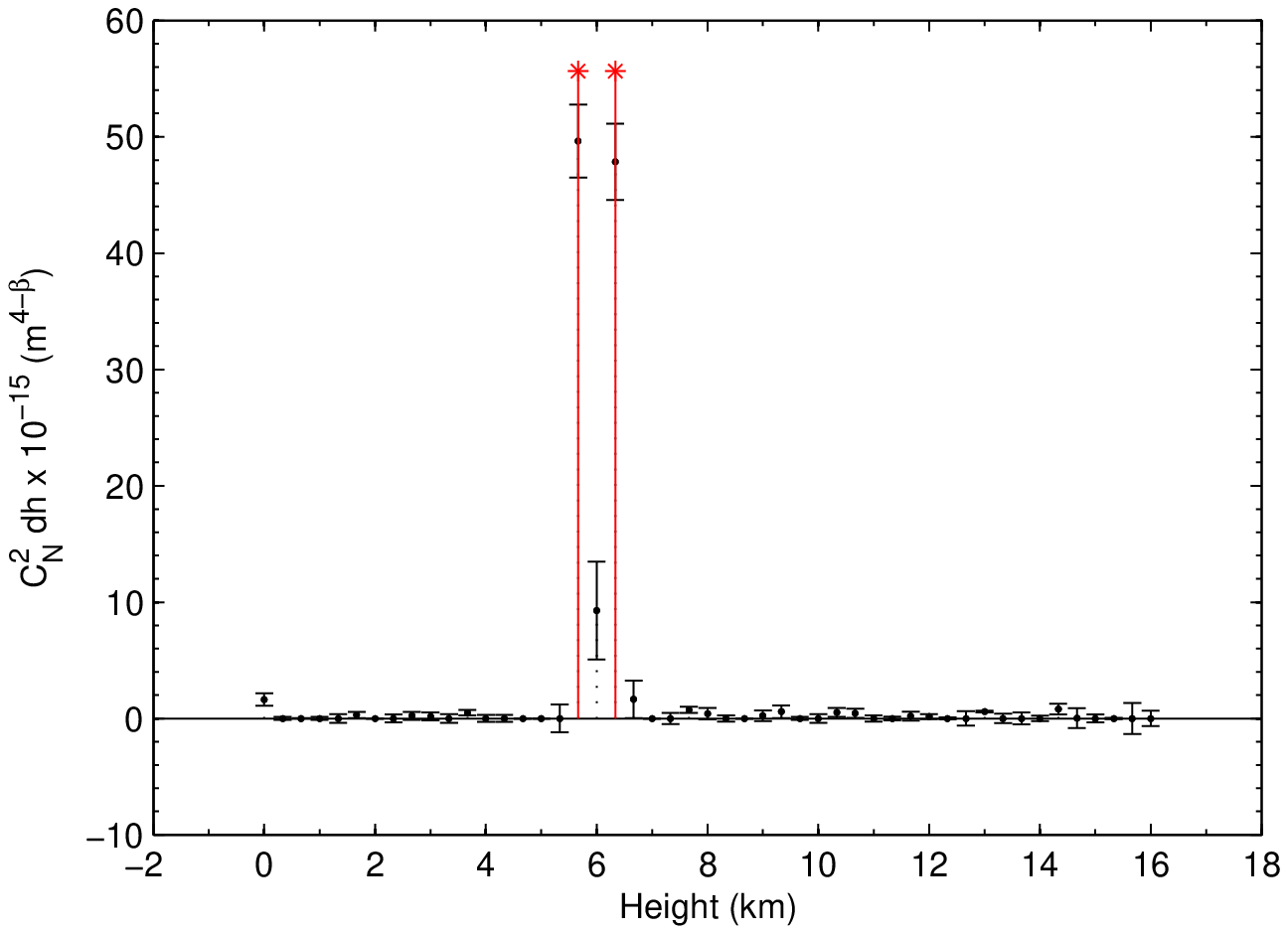}}}
      }
    \caption{Numerical simulation of Generalized SLODAR using parameters of Tab.~\ref{tab:numparams}. The objective is to fully separate two thin turbulent layers with height separation, $\Delta H=2\delta h^*$ where $\delta h^*=\delta h/N_G$ with  $\delta h = 1200~\rm{m}$ and $N_G=3$ generalized datasets. Plots (a)-(d) denote the numerical results as filled circles with error bars (black). Plot (a) shows the averaged longitudinal auto-covariance profile using star A; plot (b) shows the combined super-resolution longitudinal cross-covariance profile, $\mathbf{C}_{L}^{*'x,obs} (\delta m^*)$, using star A and star B; plot (c) shows the super-resolution $C_N^{2}(h^*)dh^*$ profile obtained by fitting super resolution kernel, $X_L^* (\Omega ^* ,\delta m^* )$, to the cross-covariance profile, plot (b), using 4000 atmospheric realizations; plot (d) is similar to plot (c) except using 2000 atmospheric realizations. Plot (a) denotes the best theoretical fit with parameters $\beta=3.63$ and $\rho_0=0.31\pm0.01~\rm{m}$ as continuous line (red); theoretical G-tilt of sub-aperture as circle (black); theoretical Z-tilt of sub-aperture as square (black). Plot (b) denotes the best theoretical fit of super-resolution $C_N^{2}(h)dh$ profile, plot (c), as continuous line (red). Plots (c) and (d) denote the modelled atmosphere with parameters $\beta=3.67$ and $\rho_0=0.3$ as stem lines with asterisks (red). The numerical simulation results shown in plots (a)-(d) confirm the validity of using the methodology outlined in Section~\ref{sec:verticalres} and illustrated in Fig.~\ref{fig:SLODAR_Generalized_Methodology}.}
    \label{fig:SLODAR_Generalized_Numerical}
  \end{center}
\end{figure}

The results of the simulation clearly separated the phase screens as illustrated in Fig.~\ref{fig:SLODAR_Generalized_Numerical}.

\begin{figure}[htbp]
  \begin{center}
    \mbox{
      \subfigure[]{\scalebox{1.0}{\includegraphics[width=0.45\textwidth]{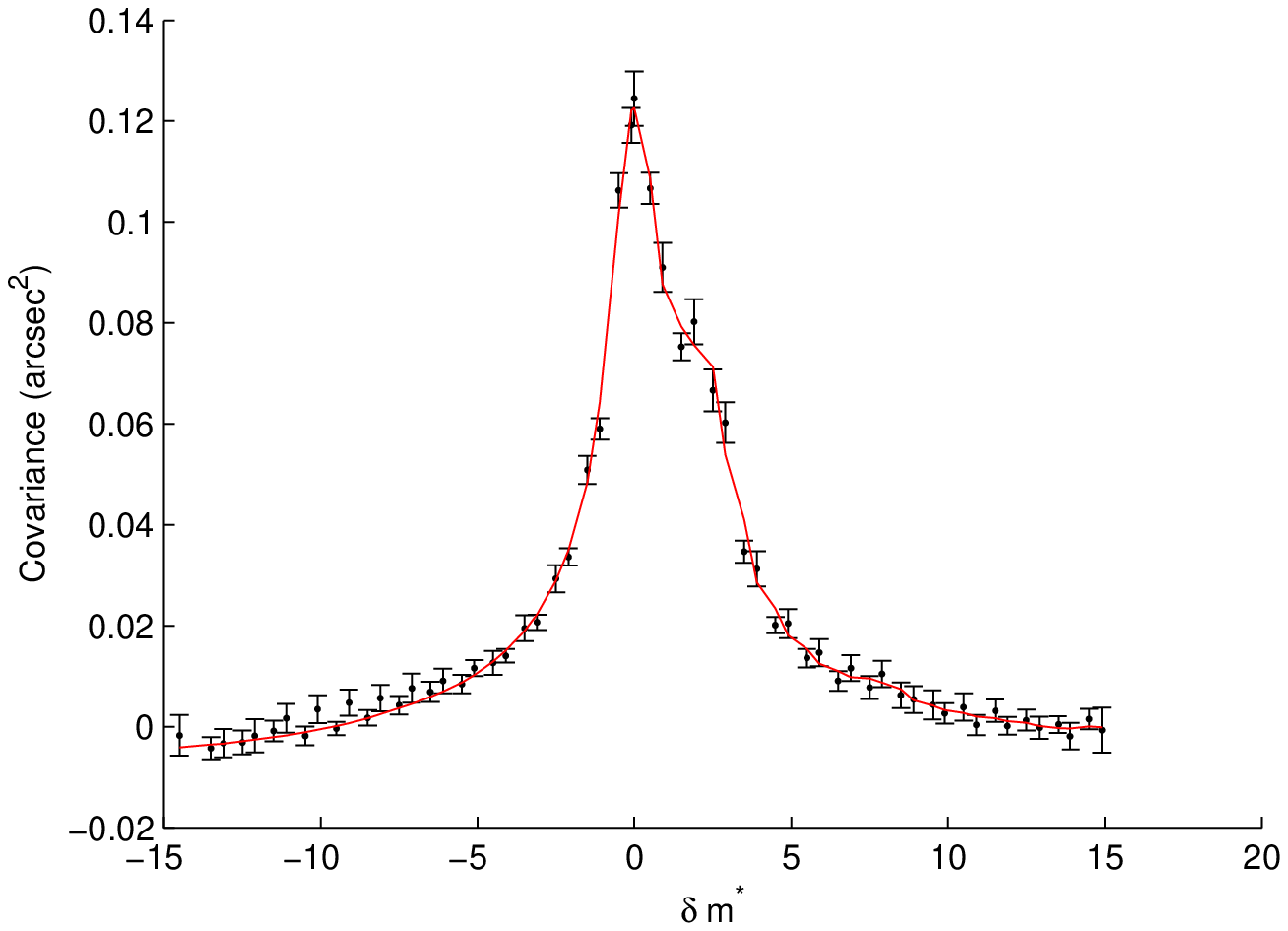}}} \quad
      \subfigure[]{\scalebox{1.0}{\includegraphics[width=0.45\textwidth]{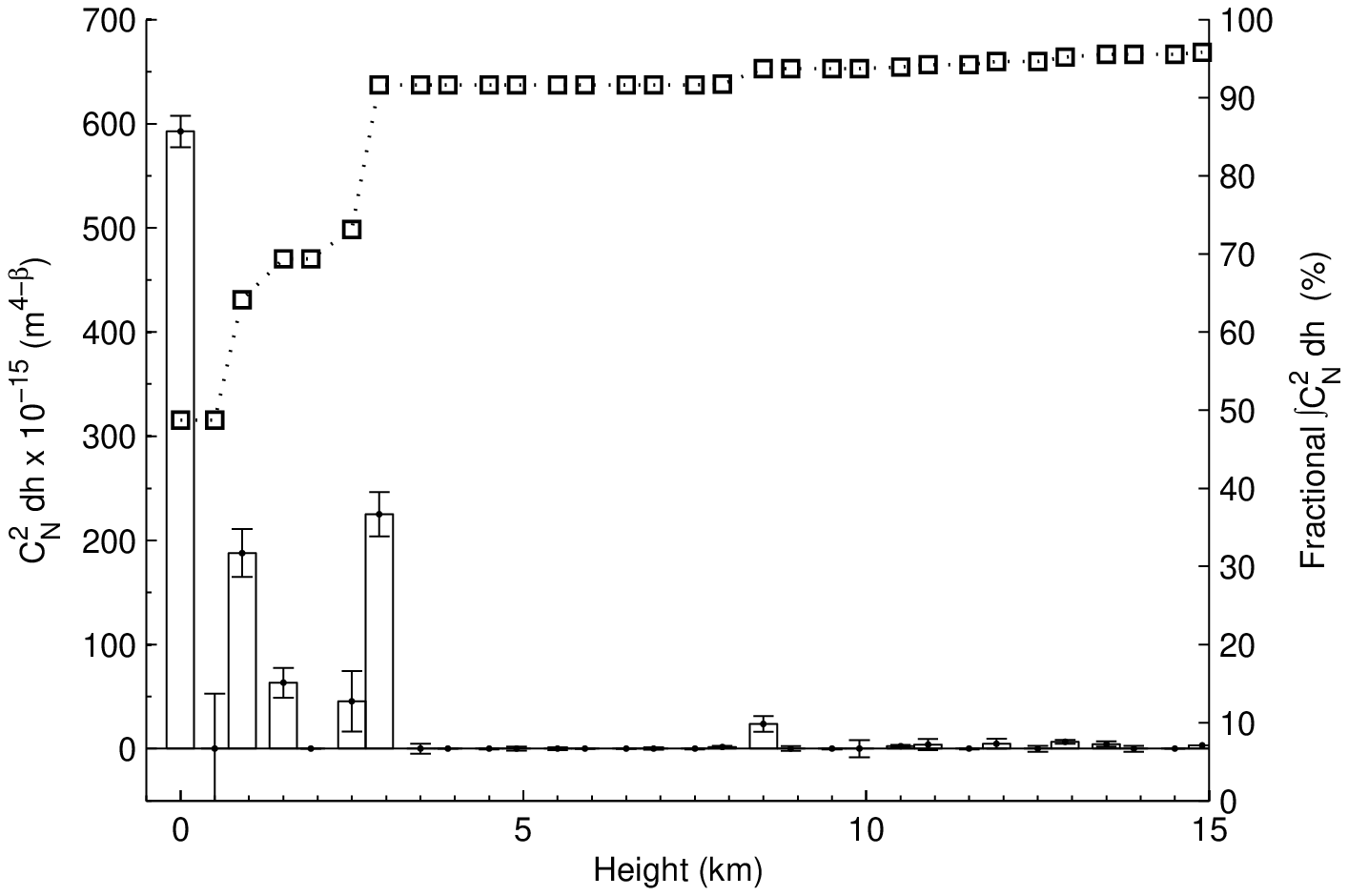}}}
      }
    \caption{Observational data of Generalized SLODAR, $N_G=2$ datasets, having fractional offsets $\eta_1=0.5$ and $\eta_2=0.9$, and nominal height resolution, $\delta h = 1102~\rm{m}$. The double star is $\alpha~\rm{Cen}$, with separation $\theta=9.44"$, observed 10:04 ($\eta_1$) \& 10:59 ($\eta_2$) 21 June 2006 (UTC) with the ANU 17x17 SLODAR instrument on the 40" telescope at SSO. The plots denote the observation results as filled circles with error bars (black). Plot (a) shows the combined super-resolution transverse cross-covariance profile, $\mathbf{C}_{T}^{*'x,obs} (\delta m^*)$, using 4000 frames for each dataset. Plot (b) shows the super-resolution $C_N^{2}(h^*)dh^*$ profile obtained by fitting super resolution kernel, $X_T^* (\Omega ^* ,\delta m^* )$, to the cross-covariance profile, shown as continuous line (red) in plot (a).  The observational data gives the seeing conditions as $\rho_0=0.095\pm0.004~\rm{m}$ and a power law of $\beta=3.15\pm0.04$. Plot (b) shows a 2x improvement in nominal resolution, $\delta h^*\sim \delta h/2$, for the $C_N^{2}(h^*)dh^*$ profile, indicating strongest turbulence is near the ground $\leq50~\rm{m}$.}
    \label{fig:SLODAR_Generalized_Observations}
  \end{center}
\end{figure}

We now apply the Generalized SLODAR methodology to observational data by combining two SLODAR datasets of the double star $\alpha~\rm{Cen}$, angular separation of 9.44", at Siding Spring Observatory.  The first dataset was captured at 10:04 21 June 2006 (UTC) with SHWFS conjugation height $h_0^*=550~\rm{m}$ ($\eta_1=0.5$) and second dataset was captured at 10:59 21 June 2006 (UTC) with SHWFS conjugation height $h_0^*=990~\rm{m}$ ($\eta_2=0.9$). The third dataset having SHWFS conjugation height $h_0=0~\rm{m}$ ($\eta_0=0$) was taken 12:43 21 June 2006 (UTC) and excluded in the analysis as the atmospheric seeing changed significantly (poor seeing) during the 1hr45mins of observing downtime. Note the fractional spacings of $\eta_1=0.5$ and $\eta_2=0.9$ are not regularly spaced but the methodology and results remain valid. Each dataset consists of 4000 frames captured at 200 fps using a fixed exposure of 2~ms with centroid sequences from each lenslet pre-processed by 1~Hz high pass FIR filter to remove mirror and dome seeing contributions from the ground-layer turbulence measurement bin. The results are shown in Fig.~\ref{fig:SLODAR_Generalized_Observations} and clearly demonstrate an improvement in height resolution by a factor two over the nominal resolution of 1100~m, providing an 'effective' resolution of 550~m. The error bars are one standard deviation calculated by dividing the dataset into 10 segments of 400 frames. As the SHWFS conjugation height $h_0=0~\rm{m}$ ($\eta_0=0$) was excluded from the analysis we added a single impulse response function for $\Omega^*=0$ to model the ground-layer.  We note that the turbulence bin for height 550~m does not register any strength. The error bar for this bin constrains the lowest turbulence to be below $\sim50~m$, as otherwise the finite width of the covariance impulse response for layers at $\sim50~m$ would cause spill-over exceeding the error bar.

\section{Conclusions}

SLODAR is a simple and valuable technique, particularly for investigation of the ground layer with inexpensive and simple equipment. We have shown that some care needs to be taken in the analysis of SLODAR data when small (few cm) sub-apertures are used in the Shack-Hartmann wavefront sensor, as Fresnel propagation effects can lead to underestimation of high layers, thereby overestimating the importance of the common strong ground layers. We have shown that pre-filtering the centroid data stream with a high-pass filter with a cut-off around 1-2 Hz can remove mirror and dome seeing providing an accurate atmospheric ground-layer measurement. We have also shown that a simple optical technique called Generalized SLODAR can yield improved vertical resolution in the ground layer at the same time as measuring the high-altitude turbulence.

\section*{Acknowledgements}

The authors would like to acknowledge R.~Johnston, C.~Harding and R.~Lane (University of Canterbury) for MATLAB source code (April 1999) to simulate a phase screen with Kolmogorov statistics using interpolative methods~\cite{Harding}. The phase screens were used as part of the numerical simulation to examine Generalized SLODAR (see Section~\ref{sec:verticalres}, Fig.~\ref{fig:SLODAR_Generalized_Numerical}).

The authors would like to thank M.~C. Britton (California Institute of Technology), the author of Arroyo~\cite{Britton}, a publicly available cross-platform C++ class software library for simulation of adaptive optic systems. The software library was used to simulate propagated and non-propagated phase screens. The phase screens were used as part of the numerical simulation to examine Fresnel propagation effects (see Section~\ref{sec:propagation}, Fig.~\ref{fig:SLODAR_Numerical_Propagation}).

The authors would like to thank P.~C. Hansen (Technical University of Denmark), the author of RegTools (Regularization Tools)~\cite{Hansen}, A publicly available MATLAB package for Analysis and Solution of Discrete Ill-Posed Problems. The regularization tools were helpful in exploring and understanding the best regularization methods to use for SLODAR.

\end{document}